\documentclass[aps,twocolumn,amsmath,amssymb,preprintnumbers]{revtex4}
\usepackage{amsmath} \usepackage{amsfonts} \usepackage{amssymb}
\usepackage{bbm}
\usepackage{epsfig}
\usepackage{graphics}
\usepackage{graphicx}
\newcommand{\be}{\begin{equation}}
\newcommand{\ee}{\end{equation}}
\newcommand{\ba}{\begin{eqnarray}}
\newcommand{\ea}{\end{eqnarray}}
\newcommand{\nn}{\nonumber}
\newcommand{\mn}{_{\mu\nu}}
\newcommand{\hmn}{_{\hat\mu\hat\nu}}
\newcommand{\imn}{^{\hat\mu\hat\nu}}

\newcommand{\ab}{_{\alpha\beta}}
\newcommand{\dz}{^{(D_2)}}
\newcommand{\de}{^{(D_1)}}
\newcommand{\M}{M^{D_1-2}}
\newcommand{\vd}{\varphi_\delta}

\begin{document}

\title[ ]{Constraints for warped branes}

\author{C. Wetterich}
\affiliation{Institut  f\"ur Theoretische Physik\\
Universit\"at Heidelberg\\
Philosophenweg 16, D-69120 Heidelberg}

\begin{abstract}
We investigate singular geometries which can be associated with warped branes in arbitrary dimensions. If the brane tension is allowed to be variable, the extremum condition for the action requires additional constraints beyond the solution of the field equations. In a higher dimensional world, such constraints arise from variations of the metric which are local in the usual four-dimensional spacetime, without changing the geometry of internal space. As a consequence, continuous families of singular solutions of the field equations, with arbitrary integration constants, are generically reduced to a discrete subset of extrema of the action, similar to regular spaces. As an example, no static extrema of the action with effective four-dimensional gravity exist for six-dimensional gravity with a cosmological constant. These findings explain why the field equations of the reduced four-dimensional theory are not consistent with arbitrary solutions of the higher dimensional field equations - consistency requires the additional constraints. The characteristic solutions for variable tension branes are non-static runaway solutions where the effective four-dimensional cosmological constant vanishes as time goes to infinity. 
\end{abstract}

\maketitle

\section{Introduction}
It is conceivable that our observed four-dimensional world corresponds to a generalized membrane - ``brane'' - embedded in a higher dimensional world \cite{ARS}, \cite{CWWB}, \cite{PR}. The presence of the brane modifies the higher dimensional geometry. One important ingredient is the warping factor, which is the analogue of the gravitational potential for a black hole solution in four dimensions. (Black holes are special cases of pointlike branes.) Higher dimensional spaces with warping have been investigated intensively in the past \cite{RSW}, \cite{CWWB}, \cite{RDW}, \cite{RS}. A continuous family of warped solutions has been found in six-dimensional gravity with a cosmological constant, depending on free integration constants \cite{RSW}. In particular, the effective four-dimensional cosmological constant can be considered as one of the integration constants, being therefore a property of the solution rather than given by the parameters of the higher dimensional theory. While the geometries corresponding to these solutions are rather strange and do not lead to an acceptable effective four-dimensional theory, similar features have been observed for warped branes in the six-dimensional Einstein-Maxwell theory \cite{CWWB}. These solutions describe codimension two singularities, similar to strings in four dimensions, and do not share some of the problems of the solutions in ref. \cite{RSW}. In modern language, these are codimension two branes \cite{GS},\cite{CS}, \cite{9A}- actually the first warped brane solutions. (The equivalence of the bulk point of view with ``non-compact'' or singular geometries and the brane point of view where a brane sits at the singularity is discussed extensively in \cite{SCW}, see also \cite{GS}.)

One may ask the question if all the warped solutions within these continuous families are equally acceptable. Since some of the integration constants are directly related to the brane tension \cite{SCW}, the issue if all integration constants are allowed depends on the status of the brane. We may distinguish between two types of branes - fixed tension branes and variable tension branes. For fixed tension branes the tension is treated as a fixed parameter of the model, while for variable tension branes the tension is a property of a particular solution and may change in the course of the cosmological evolution. Only for variable tension branes the tension has the status of an integration constant. In contrast, for fixed tension branes one of the integration constants of the most general ``bulk solution'' is fixed by the brane tension.

In our four-dimensional world we can find analogues of the two types of branes. An example for fixed tension branes are strings in grand unified theories (cosmic strings). Here the string tension can be calculated from the parameters of the microscopic theory. It should be considered as a fixed parameter for the effective theory at distances larger than the transverse size of the string. (In this effective theory the string appears as a singular object, while in the microscopic theory it is regular.) An example for variable tension branes are black holes. Here the strength of the singularity is determined by the mass of the black hole $M_{BH}$. It is a property of a particular solution, reflecting the cosmological history of a given astrophysical object. The mass of the black hole can change in the course of the history by accretion of mass. Thus the strength of the singularity should indeed be considered as an integration constant.

For four-dimensional black holes one finds a continuous family of static isotropic Schwarzschild solutions, parameterized by the integration constant $M_{BH}$. By analogy, one may therefore guess that variable tension branes in higher dimensions also lead to continuous families of solutions, with free integration constants. In this paper we argue that in general this is not the case. 

If the brane tension is variable, metric fluctuations that change the brane tension cannot be excluded. The extremum condition for the action with respect to such fluctuations leads to constraints beyond the higher dimensional field equations. These additional ``brane constraints'' exclude most of the values of the integration constants of the general solution of the field equations. The case of a continuous family of solutions similar to the black holes is an exceptional special case. The generic situation admits at most a discrete number of solutions, similar to regular spaces. In particular, the continuous family of solutions of ref. \cite{RSW} cannot be realized by variable tension branes - none is an extremum of the action with respect to arbitrary variations of the metric. Only for a fixed tension brane one particular solution becomes consistent, with integration constants chosen according to the brane tension. In this case, however, the four-dimensional cosmological constant is fixed by the brane tension as a parameter of the theory. It appears no longer as an integration constant associated with a particular solution. (We employ the wording of ``brane tension'' here in a generalized sense for the strength of the singularity. It always characterizes the singular energy momentum tensor in the ``brane picture'', even though the singularities are sometimes not really branes in a stricter sense, as for the solutions in \cite{RSW}.)

Solutions of the higher dimensional field equations are guaranteed to be extrema of the action with respect to all {\em local} field variations. Local field variations in a higher dimensional sense are restricted to a neighborhood around some point in higher dimensions. They do not change the strength of a singularity. In contrast, local field variations in a four-dimensional sense are only local around some point in four-dimensional space, but not restricted to be local in internal space as well. For example, fluctuations that do not change the shape of internal geometry, while changing the four-dimensional metric locally, are local in a four-dimensional sense, but not in a higher dimensional sense. Such variations often change the strength of the singularity. They may therefore be forbidden for fixed tension branes, while they count among the natural variations for variable tension branes. 

An effective four-dimensional theory can be obtained by expanding in fluctuations around a given higher dimensional solution, and integrating over internal space. The infinitely many ``Kaluza-Klein modes'' are usually truncated to a finite subset. Typically, the effective four-dimensional theory contains the four-dimensional metric as well as four-dimensional scalar fields, the latter corresponding to changes of size and shape of the internal geometry, that are local in a four-dimensional sense. (In addition, there may be fermions and gauge fields which are not relevant for our discussion.) An extremum of the higher dimensional action should be reflected as an extremum of the effective four-dimensional action. In the Einstein frame, where the coefficient of the four-dimensional curvature scalar is a constant, an extremum of the four-dimensional effective action has to correspond to an extremum of the effective four-dimensional scalar potential $V$. This should hold whenever the higher dimensional solution has static size and shape of internal space and static warping. An explicite computation shows that this is not the case for arbitrary integration constants of the higher dimensional solutions.

In general, the four-dimensional field equations for the infinitely many fields are more restrictive than the higher dimensional field equations. The four-dimensional field equations are the extremum conditions for the action with respect to all fluctuations which are local in a four-dimensional sense. In contrast, the higher dimensional field equations correspond to variations of fields which are local in a higher dimensional sense. The second class of fluctuations is only a subset of the first. For a singular internal geometry, solving the  higher dimensional field equations is a necessary, but not a sufficient condition for solving the four-dimensional field equations. 

As an illustrative example we may discuss the solutions of ref. \cite{RSW} for six dimensional gravity with a cosmological constant. They have two internal dimensions, $D=2$, with $SO(D)$ symmetry. The internal geometry and the warping are static, and the effective four-dimensional cosmological constant $\Lambda$ is a free integration constant. We may generalize this construction to arbitrary $D$, within $D+4$ dimensional gravity with a cosmological constant \cite{RDW}. For the effective four-dimensional theory we consider here variations of the size of internal space, $L(x)$, and variations of the scale of the $D-1$ dimensional subspace on which the $SO(D)$ symmetry acts, denoted by $A(x)$. The four-dimensional scalar field corresponding to $L(x)$ is often called the radion. The solution of the higher dimensional field equations corresponds to $L=A=1$, and we consider variations which are local in the coordinates of the usual four-dimensional space $x^\mu$. Nevertheless, an overall rescaling of the ``internal metric'' by $L$ modifies the strength of the singularity. Also variations of the four-dimensional metric change the strength of the singularity.

In sect. \ref{Einsteinspaces} we discuss the most general solutions of the higher dimensional field equations which lead to a finite coefficient of the four-dimensional curvature scalar or a finite effective reduced Planck mass $M$. Finite $M$ is required for a consistent effective four-dimensional gravity \cite{7A}. Generically, these solutions have two singularities. One is a brane, the other is a ``zerowarp'' solution for which the warping factor vanishes at the singularity. For special choices of the integration constants the brane singularity gets replaced by a regular behavior - these are the solutions discussed in refs. \cite{RSW}, \cite{RDW}. After dimensional reduction, we compute the effective four-dimensional scalar potential for $L$ and $A$ in sects. \ref{Cosmologywith}, \ref{runawaysolutions}, 
\ba\label{A}
V&=&L^{-D}A^{-\frac{D-1}{2}}
\left\{\tilde c_V+\left(2\Lambda-\frac{D+4}{D+2}\tilde c_V\right)L^{-2}\right.\nn\\
&&\left.+\frac{(D-1)(D-2)\tilde c_A}{2L^2}
\left(1-\frac{1}{A}\right)\right\}.
\ea
Here $\tilde c_V$ and $\tilde c_A$ are constants that depend on the particular higher dimensional solution around which we expand, in particular on the integration constant $\Lambda$. The constant $\tilde c_V$ is proportional to the higher dimensional cosmological constant.

An extremum of the higher dimensional action requires that $V(L,A)$ should have an extremum for $L=A=1$. Only for these values of $L$ and $A$ the higher dimensional field equations are obeyed. Inspection of eq. \eqref{A} shows, however, that for $D=2$ the potential has no extremum at all, while for $D>2$ an extremum at $L=A=1$ requires the ``brane constraints''
\be\label{B}
\Lambda=\frac{2}{D+2}\tilde c_V~,~\tilde c_A=\frac{2\tilde c_V}{(D-2)(D+2)}.
\ee
The first brane constraint fixes the effective four-dimensional cosmological constant in terms of the higher dimensional cosmological constant. (The specific higher dimensional solution obeying the condition to \eqref{B} turns out to be actually a maximum of $V$  for a positive higher dimensional cosmological constant, $\tilde c_V>0$.) More generally, the extremum condition for $V$ leaves at most a discrete number of solutions out of the continuous family of higher dimensional solutions. This is similar to the situation for regular spaces, where the regularity condition restricts the choice of the integration constants. In fact, higher dimensional solutions with regular spaces are automatically extrema of the effective four-dimensional potential $V$. 

The generic solutions of refs. \cite{RSW,RDW} correspond  to extrema of the higher dimensional action only for fixed tension branes. In this case variations of the overall size of internal space corresponding to $L(x)$ are not allowed since they change the brane tension. There is then no need to impose the constraint \eqref{B} anymore. As mentioned before, for a fixed tension brane the cosmological constant $\Lambda$ is fixed in terms of the brane tension. For $D>2$ specific higher dimensional solutions can also be extrema for variable tension branes. These are the ones obeying the brane constraint \eqref{B}, and $\Lambda$ is now fixed by this constraint. In conclusion, both for fixed and variable tension branes there are no continuous families of static  solutions with arbitrary integration constant $\Lambda$. For fixed tension branes $\Lambda$ is fixed by the boundary conditions at the singularity, while for variable tension branes $\Lambda$ is fixed by the brane constraint \eqref{B} which is needed to guarantee an extremum of the action. 

Even for fixed tension branes additional constraints of the type \eqref{B} arise if the field content of the higher dimensional theory is extended beyond the graviton. For example, if we add in the higher dimensional model a scalar field, the most general solution of the field equations consistent with our symmetry exhibits two additional free integration constants. Even for a fixed brane tension, the four-dimensional cosmological constant $\Lambda$ will depend on these additional integration constants. It will appear again as an integration constant rather than being fixed by the brane tension. However, there are also new brane constraints of the type \eqref{B}, connected to the extremum condition of the higher dimensional action with respect to arbitrary variations of the scalar field. Typically, these constraints select again only a discrete subset of solutions, such that $\Lambda$ remains no longer a free integration constant.

For variable tension branes, static solutions are difficult to be realized because the brane constraint may have no appropriate stable solutions. This observation has interesting consequences for cosmology. The characteristic solutions corresponding to extrema of the action turn out to be non-static runaway solutions. If the brane tension is allowed to change, it generally does so in the course of the cosmological solution. For the potential \eqref{A} one finds that $L$ and $A$ move to infinity or zero, thereby changing the strength of the singularity. A characteristic runaway solution with $L\to\infty$ leads to an asymptotically vanishing four-dimensional effective potential for $t\to\infty$, similar to quintessence cosmologies \cite{CWQ}. The fact that the potential goes to zero and not to a constant is no accident \cite{CWCC}. This property is due to geometry and will not be affected by quantum fluctuations. Thus the question why the cosmological constant vanishes asymptotically even in the presence of quantum fluctuations finds a natural solution, incidentally demonstrating that too naive arguments about the generic constribution of quantum fluctuations to the cosmological constant are not correct \cite{CWCC}. Nevertheless, runaway solutions lead to another potential problem. It has to be explained why the couplings of the standard model of particle physics change only very little (at least in the present cosmological epoch) in an environment of evolving geometry and warping of internal space. Several suggestions have been made for possible stabilization mechanisms for the couplings \cite{CWCC}, but this is not the subject of the present paper.

This paper is organized as follows. In sect. \ref{feextremumconditions} we display the field equations and the extremum conditions for $d$-dimensional gravity with a scalar field. The most general ansatz consistent with $SO(D)$ symmetry and maximal Poincare, de Sitter or anti-de Sitter symmetry in $D_1$ dimensions $(d=D_1+D)$ is displayed in sect. \ref{warpedsolutions}, where we also discuss the field equations for this ansatz. (For a realistic world one would need $D_1=4$.) In sect. \ref{Einsteinspaces} we specialize to pure gravity without a $d$-dimensional scalar field. We discuss the most general static solution consistent with an effective $D_1$-dimensional gravity. 

The special case of a vanishing $d$-dimensional cosmological constant and flat $D_1$-dimensional space is discussed in detail in the appendix. This covers the geometry of flat branes embedded in flat space for arbitrary dimensions of the branes, $D_1$, and arbitrary dimension of space, $d$. We also investigate there another class of solutions with finite volume and finite $D_1$-dimensional gravitational constant. The discussion of sect. \ref{Einsteinspaces} is extended to a $d$-dimensional scalar field in sect. \ref{Models}.

In sect. \ref{Dimesionalreduction} we perform for these higher dimensional solutions the dimensional reduction to an effective $D_1$-dimensional theory. Even if we retain only the $D_1$-dimensional metric, the effective $D_1$-dimensional Einstein equation clashes with the general $d$-dimensional solution. A simultaneous solution to the $d$-dimensional field equation and the effective field equation for the $D_1$-dimensional metric requires the brane constraint \eqref{B}. In sect. \ref{fieldvariations} we discuss how this apparent clash can be understood by a discussion of arbitrary variations of the $d$-dimensional fields. In the presence of singularities the $d$-dimensional field equations are not sufficient in order to guarantee an extremum of the action. We turn to the cosmology of variable tension branes in sect. \ref{Cosmologywith}. Starting at some time $t_0$ with a higher dimensional solution, we see that the cosmological evolution for $t>t_0$ corresponds to a static flat or (anti-) de Sitter space only if the constraint \eqref{B} is met. Otherwise, the internal geometry does not remain static. The typical solutions are rather runaway solutions where the effective four-dimensional constant asymptotically relaxes to zero, as for quintessence cosmologies \cite{CWQ}. We discuss these runaway solutions in sect. \ref{runawaysolutions}. Finally, we present our conclusions and a discussion of the possible impact of our findings for the cosmological constant problem in sect. \ref{X}.

\section{Field equations and extremum conditions}
\label{feextremumconditions}
We start with the $d$-dimensional action for the metric $\tilde g\hmn$ and a scalar field $\delta$
\be\label{c00}
S=M^{d-2}_d\int\tilde g^{1/2}
\left\{\frac12\tilde R-\frac12\partial^{\hat\rho}\delta\partial_{\hat\rho}\delta-\tilde V(\delta)\right\}.
\ee
Here $M^{d-2}_d\tilde V(\delta)$ is the higher dimensional scalar potential. In the absence of a higher dimensional scalar field a constant $\tilde V$ corresponds to the $d$-dimensional cosmological constant. We denote $d$-dimensional indices by $\hat\mu,\hat\nu,\hat\rho$, whereas $D_1$-dimensional indices will be $\mu,\nu,\rho$. 

The field equations for the metric read 
\be\label{c01}
\tilde R_{\hat\mu\hat\nu}-\frac12\tilde R\tilde g_{\hat\mu\hat\nu}=
\partial_{\hat \mu} \delta\partial_{\hat \nu} \delta-\frac12\partial^{\hat\rho}\delta\partial_{\hat\rho}
\delta\tilde g_{\hat\mu\hat\nu}-\tilde V\tilde g_{\hat\mu\hat\nu},
\ee
and the scalar field obeys
\be\label{c02}
\hat D^2\delta=\frac{\partial\tilde V}{\partial\delta}.
\ee
Here $\hat D^2=D^{\hat\rho}D_{\hat\rho}$ is the covariant Laplacian. By contracting eq. \eqref{c01} one obtains
\be\label{c03}
\tilde R=\partial^{\hat\rho}\delta\partial_{\hat\rho}\delta+\frac{2d}{d-2}\tilde V.
\ee
For solutions of the field equations the action therefore becomes
\be\label{c04}
S=\frac{2}{d-2}M^{d-2}_d\int\tilde g^{1/2}\tilde V.
\ee

In the presence of singularities the extremum condition for $S$ goes beyond the field equations. We may first discuss a general variation of the scalar field $\delta(\hat x)\to\delta(\hat x)+\epsilon(\hat x)$, with $\epsilon$ infinitesimal and $\hat x$ the $d$-dimensional coordinates. The variation of the action \eqref{c00} reads
\ba\label{92}
\delta S&=&-M^{d-2}_d\int \tilde g^{1/2}
\left\{\partial^{\hat\rho}\delta\partial_{\hat\rho}\epsilon+
\frac{\partial\tilde V}{\partial\delta}\epsilon\right\}\nn\\
&=&M^{d-2}_d\int\tilde g^{1/2}\left(\hat D^2\delta-\frac{\partial\tilde V}{\partial\delta}\right)\epsilon\nn\\
&&-M^{d-2}_d\int\partial_{\hat\rho}
\{\tilde g^{1/2}\epsilon\partial^{\hat\rho}\delta\}.
\ea
The first term in the second equation \eqref{92} vanishes by virtue of the field equation \eqref{c02}. The second term is a boundary term. It vanishes automatically for compact spaces  or for variations $\epsilon$ with a compact support. In the presence of singularities, there are possible variations $\epsilon$ for which the boundary term does not vanish automatically. The extremum condition $\delta S=0$ then requires additional constraints beyond the field equations - the brane constraints. 

A similar boundary term for the metric will be the origin of the constraint \eqref{B}. For a variation $\tilde g_{\hat\mu\hat\nu}(\hat x)\to\tilde g\hmn(\hat x)+\tilde h\hmn(\hat x)$ the extremum condition for the action requires in addition to the gravitational field equation \eqref{c01} the vanishing of a ``boundary term''
\be\label{KXA}
\int_{\hat x}\partial_{\hat\nu}(\tilde g^{1/2}K^{\hat\nu})=0,
\ee
with
\ba\label{KXB}
&&K^{\hat\nu}=(\tilde g\imn\tilde g^{\hat\rho\hat\sigma}-\tilde g^{\hat\mu\hat\rho}\tilde g^{\hat\nu\hat\sigma})\partial_{\hat\mu}\tilde h_{\hat\rho\hat\sigma}\\
&&-\frac12(\tilde g\imn\tilde g^{\hat\rho\hat\tau}-\tilde g^{\hat\mu\hat\rho}\tilde g^{\hat\nu\hat\tau})
(\partial_{\hat\mu}\tilde g_{\hat\rho\hat\sigma}+\partial_{\hat\rho}\tilde g_{\hat\mu\hat\sigma}-\partial_{\hat\sigma}\tilde g_{\hat\mu\hat\rho})
\tilde g^{\hat\sigma\hat\lambda}
\tilde h_{\hat\lambda\hat\tau}.\nn
\ea

\section{Warped solutions}
\label{warpedsolutions}
We discuss possible solutions with the ansatz
\be\label{A1}
\tilde g_{\hat\mu\hat\nu}(x,y,z)=\left(\begin{array}{ccccc}
\sigma(z)g^{(D_1)}\mn(x)&,&0&,&0\\
0&,&\rho(z)g^{(D_2)}_{\alpha\beta}(y)&,&0\\
0&,&0&,&1
\end{array}
\right)
\ee
Here the $D_2$ coordinates $y^\alpha$ describe a $D_2$-dimensional Einstein space with Ricci tensor  $R^{(D_2)}_{\alpha\beta}(y)=\Lambda_2g^{(D_2)}_{\alpha\beta}(y)$, and similar for the $D_1$ coordinates $x^\mu$, with $R^{(D_1)}\mn=\Lambda_1g^{(D_1)}\mn$. For $D_1=4$ we may identify $x^\mu$ with time and the usual three coordinates of the ``large dimensions''. The coordinates $y^\alpha$ and $z$ describe the $D=D_2+1$ dimensional internal space. The internal geometry is given by $\sigma(z),\rho(z)$ and $g^{(D_2)}_{\alpha\beta}(y)$ and therefore time-independent. The $D_1$ dimensional space is either static as well, for $\Lambda_1=0$, or it has the maximal space-time symmetry corresponding to de Sitter space $(\Lambda_1>0)$ or anti de Sitter space $(\Lambda_1<0)$. In a general sense we will refer to these solutions as static solutions. (For $\Lambda_1\neq 0$ the time translation symmetry is replace by the analogous isometry of (anti) de Sitter space.)

For definiteness, we may choose $g^{(D_2)}_{\alpha\beta}(y)$ as parameterizing a $D_2$-dimensional unit sphere, such that $\Lambda_2=D_2-1$ for $D_2>1$, and $g^{(D_2)}_{\alpha\beta}(y)=1$ for $D_2=1$. Our ansatz has then the isometry $SO(D_2+1)$ and describes the most general static metric with $SO(D_2+1)$ symmetry. The length unit is set by the radius of the $D_2$-dimensional sphere. This fixes the units for $M_d$ and $\tilde V$ in eq. \eqref{c00}. In particular, we can choose units such that $M_d=1$. The functions $\sigma$ and $\rho$ are dimensionless, while $\Lambda_1,\Lambda_2$ and $\tilde V$ have dimension mass $^2$. We will generalize the  discussion to arbitrary $\Lambda_2$ for $D_2>1$, while $\Lambda_2=0$ for $D_2=1$. Since we are free to choose units for the coordinates $y$ we can take $\Lambda_2=\pm(D_2-1)$ or $\Lambda_2=0$. Alternatively, we can also use for $y$ dimensionless angular variables, such that $\rho$ has dimension mass $^{-2}$ and $\Lambda_2$ is dimensionless. This is particularly appropriate for $D_2=1$, with periodic $y$ and $0\leq y\leq 2\pi$. 

The $\mu-\nu$-component of the field equation for the metric reads \cite{RDW}
\ba
&&2\tilde R\mn -\tilde R\tilde g\mn=\Big\{
-(D_1-2)\Lambda_1\sigma^{-1}-D_2\Lambda_2\rho^{-1}\nn\\
&&+(D_1-1)\frac{\sigma''}{\sigma}+\frac14(D_1-1)(D_1-4)
\left(\frac{\sigma'}{\sigma}\right)^2\nn\\\label{c1}
&&+\frac12(D_1-1)D_2\frac{\rho'}{\rho}\frac{\sigma'}{\sigma}+D_2\frac{\rho''}{\rho}\\
&&+\frac14D_2(D_2-3)
\left(\frac{\rho'}{\rho}\right)^2\Big\}
\tilde g\mn=-(2\tilde V+\delta'^2)\tilde g\mn,\nn
\ea
and similar for the $\alpha-\beta$-component
\ba
&&2\tilde R_{\alpha\beta}-\tilde R\tilde g_{\alpha\beta}=
\Big\{-(D_2-2)\Lambda_2\rho^{-1}-D_1\Lambda_1\sigma^{-1}\nn\\
&&+(D_2-1)\frac{\rho''}{\rho}+\frac14(D_2-1)(D_2-4)
\left(\frac{\rho'}{\rho}\right)^2\nn\\
&&+\frac12(D_2-1)D_1\frac{\rho'}{\rho}\frac{\sigma'}{\sigma}+D_1\frac{\sigma''}{\sigma}\label{c2}\\
&&+\frac14 D_1(D_1-3)
\left(\frac{\sigma'}{\sigma}\right)^2\Big\}\tilde g_{\alpha\beta}
=-(2\tilde V+\delta'^2)\tilde g_{\alpha\beta},\nn
\ea
and the $z-z$-component
\ba
&&2\tilde R_{zz}-\tilde R \tilde g_{zz}=-D_1\Lambda_1\sigma^{-1}-D_2\Lambda_2\rho^{-1}\nn\\
&&+\frac12 D_1D_2\frac{\rho'}{\rho}\frac{\sigma'}{\sigma}+\frac14 D_1(D_1-1)\left(\frac{\sigma'}{\sigma}\right)^2\nn\\
&&+\frac14 D_2(D_2-1)\left(\frac{\rho'}{\rho}\right)^2
=-(2\tilde V-\delta'^2).\label{c3}
\ea
Correspondingly, the curvature scalar obeys
\ba\label{C3A}
\tilde R&=&D_1\Lambda_1\sigma^{-1}+D_2\Lambda_2\rho^{-1}-D_1
\frac{\sigma''}{\sigma}-D_2\frac{\rho''}{\rho}\nn\\
&&-\frac14 D_1(D_1-3)\left(\frac{\sigma'}{\sigma}\right)^2-\frac14 D_2(D_2-3)
\left(\frac{\rho'}{\rho}\right)^2\nn\\
&&-\frac12 D_1D_2\frac{\rho'}{\rho}\frac{\sigma'}{\sigma}\nn\\
&=&\delta'^2+\frac{2(D_1+D_2+1)}{D_1+D_2-1}\tilde V.
\ea
All other components vanish. 

Using
\be\label{c3A}
U=\frac{\sigma'}{\sigma}\quad,\quad W=\frac{\rho'}{\rho},
\ee
and
\be\label{c4}
F=2\tilde V-\delta'^2-D_1\Lambda_1\sigma^{-1}-D_2\Lambda_2\rho^{-1},
\ee
we can express $W$ in terms of $\rho,\sigma$ and $U$ by employing eq. \eqref{c3} for $D_2>1$,
\ba\label{c5}
&&W=\frac{1}{D_2(D_2-1)}\Big\{-D_1D_2U\\
&&+\epsilon\sqrt{D_1D_2(D_1+D_2-1)U^2-4D_2(D_2-1)F}\Big\}.\nn
\ea
Here $\epsilon=\pm 1$ corresponds to the two solutions of the quadratic equation \eqref{c3}. For $D_2=1$ the expression for $W$ simplifies and reads for $U\neq 0$
\be\label{c5A}
W=-\frac{D_1-1}{2}U-\frac{2F}{D_1U}.
\ee
Suitable linear combinations of eqs. \eqref{c1}-\eqref{c3} yield
\be\label{c6}
U'=-\frac{D_1}{2}U^2-\frac{D_2}{2}UW+\frac{2\Lambda_1}{\sigma}-\frac{4}{D_1+D_2-1}\tilde V,
\ee
and
\be\label{c7}
W'=-\frac{D_2}{2}W^2-\frac{D_1}{2}UW+\frac{2\Lambda_2}{\rho}-\frac{4}{D_1+D_2-1}\tilde V.
\ee

In addition, we have the scalar field equation
\be\label{c8}
\delta''+\left(\frac{D_1}{2}\frac{\sigma'}{\sigma}+\frac{D_2}{2}\frac{\rho'}{\rho}\right)
\delta'-\frac{\partial\tilde V}{\partial\delta}=0.
\ee
The derivative of eq. \eqref{c5}, combined with eqs. \eqref{c6} and \eqref{c8} yields eq. \eqref{c7}, which is therefore not independent. This holds provided $(D_2-1)W+D_1U\neq 0$.

For a numerical solution one may use the two second order differential equations \eqref{c6}, \eqref{c8} for $\sigma$ and $\delta$, supplemented by $\rho'=W\rho$. For $D_2>1$ the function $W(\rho,\sigma, U)$ is given by eq. \eqref{c5}, while for $D_2=1$ eq. \eqref{c5A} applies. This system is equivalent to five nonlinear first order differential equations. The five initial conditions will lead to the continuous family of solutions characterized by integration constants. For $D_2>1$ the initial conditions for $\rho,\sigma,U$ have to be chosen such that the argument of the square root of eq. \eqref{c5} is positive or zero. Alternatively, one may use eqs. \eqref{c6}-\eqref{c8}, corresponding to a system of six first order differential equations. In this case the initial conditions are not independent, however. Eq. \eqref{c3} relates the allowed initial values. In the case of pure higher dimensional gravity the scalar field equation \eqref{c5} is absent and the number of integration constants gets reduced by two.

The characteristic properties of the solution can often be understood in terms of the warping factor $\sigma(z)$. For $D_1=1$ this is the analogue of the gravitational potential $B(r)$ for the (generalized) Schwarzschild solution for black holes.
With $s=\ln(\sigma/\sigma_0),~U=s'$, eq. \eqref{c6} takes for $D_2>1$ the explicite form
\be\label{c9}
s''-\frac{D_1}{2(D_2-1)}s'^2+
\frac{\epsilon s'K}{2(D_2-1)}+\frac{\partial V_s}{\partial s}=0,
\ee
where
\ba\label{c10}
\frac{\partial V_s}{\partial s}&=&-2\tilde \Lambda_1 e^{-s}+
\frac{4}{D_1+D_2-1}\tilde V(\delta),\\
K&=&\sqrt{D_1D_2(D_1+D_2-1)s'^2-4D_2(D_2-1)F},\nn
\ea
and $\tilde\Lambda_1=\Lambda_1/\sigma_0$. This has an analogue of a particle moving in a potential
\be\label{c11}
V_s=2\tilde\Lambda_1 e^{-s}+\frac{4}{D_1+D_2-1}\tilde Vs,
\ee
with time replaced by $z$. However, the form of the ``damping terms'' involving $s'$ is rather complicated.

Under a reflection $z\to -z,~s'\to -s'$ the sign of the square root is reversed, $\epsilon\to-\epsilon$. For a positive sign  $\epsilon=1$ the effect of the terms involving $s'$ is the following. For $s'<0$ they tend to increase $s'$ as $z$ increases. (The contribution to $s''$ is positive.) This has a damping effect, since $|s'|$ gets reduced. For $s'>0$ the term $\sim \epsilon$ often dominates. Then the effect is a relative decrease of $s'$ for increasing $z$ and therefore again a damping. If for decreasing $z$ we switch the sign of $\epsilon$ we again find a damping. This contrasts to the usual damping terms in mechanics which are linear in $s'$. In our case, the damping in both directions of $z$ reflects the invariance of our system of equations under the reflection $z\to -z$. Finally, if we switch the sign $\epsilon$ without changing the direction of $z$ the damping typically turns into antidamping. Via $F$ the strength of the damping / antidamping depends on $\rho,\sigma,\tilde V$ and $\delta'^2$. 

For $D_2=1$ the equation for $s$ is simpler
\be\label{C11A}
s''+\frac{D_1+1}{4}s'^2+\frac12 \frac{\partial V_s}{\partial s}=0.
\ee
The evolution shows damping for increasing $z$ and $s'>0$, or for decreasing $z$ and $s'<0$. The opposite cases, increasing $z$ and $s'<0$ or decreasing $z$ and $s'>0$, are characterized by antidamping. The damping term can be absorbed by a variable change
\be\label{19A}
t=e^{\frac{D_1+1}{4}s}
=\left(\frac{\sigma}{\sigma_0}\right)^{\frac{D_1+1}{4}},
\ee
with
\be\label{19B}
t''=-\frac{D_1+1}{8}t\frac{\partial V_s}{\partial s}=-\frac{\partial V_t}{\partial t}
\ee
and 
\be\label{19C}
V_t=\frac{D_1+1}{4D_1}\tilde Vt^2-
\frac{(D_1+1)^2}{8(D_1-1)}\tilde\Lambda_1 t^{\frac{2(D_1-1)}{D_1+1}}.
\ee
This map to an undamped motion of a particle  in a potential $V_t$ makes a classification of the possible solutions particularly simple \cite{RSW}, \cite{SCW}. For $D_1=4$ the potential
\be\label{A6a}
V_t=\frac{5}{16}\tilde Vt^2-\frac{25}{24}\tilde \Lambda_1t^{6/5}
\ee
has a minimum if $\tilde V>0,\tilde \Lambda_1>0$, located at
\be\label{A6b}
t_{min}=\left(\frac{2\tilde\Lambda_1}{\tilde V}\right)^{5/4},
\ee
corresponding to $\sigma_{min}=2\Lambda_1/\tilde V$. Typical solutions are part of an oscillation around $t_{min}$, with a singularity $\sigma\to 0$ if $t\to 0$. For such a solution $\rho(z)$ can be computed from eq. \eqref{c5A} and singularities may occur if $\rho\to 0$ or $\rho\to\infty$.

\section{Higher dimensional Einstein spaces}
\label{Einsteinspaces}
In this section we discuss the case of  higher dimensional gravity with a cosmological constant. Thus $\tilde V$ is constant and no scalar field is present. (In our system of equations one sets $\delta'=0,\partial\tilde V/\partial\delta=0$.) Warped solutions of this type have been discussed in the original paper by Rubakov and Shaposhnikov \cite{RSW} for $D_1=4,~D_2=1,~\Lambda_2=0$, and for general $D_1,~D_2,\Lambda_1,~\Lambda_2$ by \cite{RDW}. We concentrate first on a positive cosmological constant, $\tilde V>0$. 
We will find generic solutions with two singularities, and special choices of the integration constants with only one singularity or no singularity.

For  $\Lambda_1>0$ the potential $V_s$ has a minimum at $s=s_E$,
\be\label{c12}
s_E=\ln\left(\frac{D_1+D_2-1}{2}
\frac{\tilde \Lambda_1}{\tilde V}\right),
\ee
corresponding to 
\be\label{c12A}
\sigma_E=\frac{D_1+D_2-1}{2}\frac{\Lambda_1}{\tilde V}.
\ee
As $z$ increases, one may suspect that $\sigma(z)$ approaches $\sigma_E$ and settles there. We will see that this is not necessarily the case. For example, $\rho$ may diverge or approach zero before $\sigma_E$ is reached. Nevertheless,  for $s(z)=s_E$ eq. \eqref{c9} is obeyed for arbitrary $\rho$. Indeed, $U=0$ solves eq. \eqref{c6} for constant $\sigma=\sigma_E$, and the constant warping factor implies a direct product geometry for $D$-dimensional internal space and $D_1$-dimensional large space.

\begin{itemize}
\item[]{\bf a)\quad  codimension two singularities}
\end{itemize}

Let us first discuss the case $D_2=1,~\Lambda_2=0,~D_1=4$ where eq. \eqref{c3} takes the simplified form
\be\label{x1}
2\tilde V-\frac{4\Lambda_1}{\sigma}+2\frac{\rho'}{\rho}\frac{\sigma'}{\sigma}+3
\left(\frac{\sigma'}{\sigma}\right)^2=0.
\ee
The solutions have been classified and discussed extensively in \cite{RSW}, \cite{SCW}, and we discuss here some aspects that can be generalized to more complicated situations. Consider a fixed point of the $SO(2)$ isometry at $z=0$ which is regular, as the pole of a two-sphere. This requires
\be\label{x2}
\lim_{z\to 0}\rho(z)=z^2~,~\lim_{z\to 0}\sigma(z)=\sigma_0.
\ee
The divergence of $W=\rho'/\rho=2/z$ requires that $\sigma'$ must vanish for $z\to 0$, and we make the ansatz
\be\label{x3}
U=\frac{\sigma'}{\sigma}=uz+...
\ee
Then for $z\to 0$ eq. \eqref{x1} yields the relation
\be\label{x4}
u=\frac{\Lambda_1}{\sigma_0}-\frac12\tilde V.
\ee
Without loss of generality we may take $\sigma_0=1$. The ``initial conditions'' or ``integration constants'' are all fixed by the regularity condition \eqref{x2} and $\Lambda_1$. Numerically, one may take some very small $z_0$ and use the initial conditions $\rho(z_0)=z^2_0,~\sigma(z_0)=1,~U(z_0)=uz_0$. 

Solving the differential equations for larger $z$ results in a singularity for $z\to \bar z$, where
\be\label{x5}
\sigma=\bar\sigma(\bar z-z)^{-\eta}~,~\rho=\bar\rho(\bar z-z)^\gamma~,~\eta=-\frac45~,~\gamma=-\frac65.
\ee
This singular solution exists for $\Lambda_1=0$, but also for positive and negative $\tilde\Lambda_1<3\tilde V/10$. The ''internal volume'', more precisely the volume of the two dimensional hypersurface for constant $x^\mu$, 
\be\label{x6}
\Omega_2=2\pi\int^{\bar z}_0dz\rho^{1/2}\sigma^2,
\ee
is finite, since one has $\rho^{1/2}\sigma^2\sim(\bar z-z)$ close to the singularity. Similarly, the effective gravitational constant after dimensional reduction is proportional to
\be\label{x7}
c_R=\int^{\bar z}_0d z\rho^{1/2}\sigma
\ee
and remains finite. The solution with boundary conditions \eqref{x2}, \eqref{x5} is the Rubakov-Shaposhnikov solution \cite{RSW}. The singularity has the geometry of a six-dimensional black hole, with time coordinate replaced by $y$ \cite{SCW}. 
Perhaps the most striking feature of this singularity is the vanishing of the warp factor, $\sigma(z\to\bar z)\to 0$. We will call singularities with this property ``zerowarps''. Even arbitrarily large distances on the four-dimensional space, as measured with $g^{(4)}_{\mu\nu}$, shrink to zero $d$-dimensional distances as $\bar z$ is approached. In contrast to branes, where $\rho$ shrinks to zero at the singularity \eqref{x5}, the zerowarp singularity has a diverging $\rho$.

For $\Lambda_1>0$ there is a notable special case, namely for $u=0$ or $\Lambda_1=\tilde V/2$. This special case corresponds to the minimum of the potential $V_s(s)$ in eq. \eqref{c9}. It implies a constant $\sigma(z)=\sigma_0$. (More precisely, $\sigma=const.$ follows from eq. \eqref{c9} if for some $z_0$ one has $\sigma(z_0)=\sigma_E~,~\sigma'(z_0)=0$.) The geometry is therefore a direct product of a four-dimensional de Sitter space and a two dimensional space. The $z$-dependence of $\rho$ follows from eq. \eqref{c7}
\be\label{x8}
W'+\frac12 W^2+\tilde V=0.
\ee
This equation is invariant under a rescaling of $\rho\to A_\rho\rho$, such that solutions exist with an arbitrary overall normalization of $\rho$. (This feature is particular for vanishing $\Lambda_2=0$.)

The general solution 
\ba\label{x9}
W&=&\sqrt{2\tilde V}tg \left( \frac\pi2-\sqrt{\frac{\tilde V}{2}}z\right),\nn\\
\rho&=&\frac{2\rho_0}{\tilde V}\sin^2
\left(\sqrt{\frac{\tilde V}{2}}z\right),
\ea
behaves for $z\to 0$ as
\be\label{x10}
\rho(z\to 0)=\rho_0z^2.
\ee
For $\rho_0=1$ it describes a sphere, whereas for $\rho_0\neq 1$ one has a conical singularity, with deficit angle 
\be\label{35A}
\Delta=2\pi(1-\sqrt{\rho_0)}.
\ee
A similar singularity occurs for $\bar z=\pi\sqrt{2/\tilde V}$, and we note the reflection symmetry around the point $z_{max}=\pi/\sqrt{2\tilde V}$.  This may suggest the existence of a direct product solution with a ``football shaped'' internal space \cite{9A}. However, for $U=0$ the three equations \eqref{c1}-\eqref{c3} are no longer linearly dependent. Inserting the direct product ansatz into eq. \eqref{c1} yields the condition $\Lambda_1=\tilde V$ (for $\sigma_0=1$). This is compatible with eq. \eqref{c12A} ($\Lambda_1=\tilde V/2$) only for $\tilde V=0,\Lambda_1=0$. Then the volume of internal space and $c_R$ do not remain finite. Football shaped geometries for $D_2=1$ can only be realized with additional ingredients, as for the six-dimensional Einstein-Maxwell theory \cite{RSS}. Exact time dependent cosmological solutions for such football shaped geometries have been discussed in \cite{SWCOS}.

The solutions discussed so far require special initial conditions or a special value of $\Lambda_1$. For generic initial conditions and $\Lambda_1$ we find solutions with two singularities. For $\tilde \Lambda_1<3\tilde V/10$ one is a zerowarp of the type of eq. \eqref{x5}, the other is a conical singularity. For a general conical singularity at $z=0$ the behavior of $\rho$ is given by eq. \eqref{x10} and for $\sigma$
one has
\ba\label{x11}
\sigma(z\to 0)&=&\sigma_0\left(1+\frac{u}{2}z^2\right),\nn\\
U(z\to 0)&=&uz.
\ea
From eq. \eqref{x1} or \eqref{c6} one infers the relation \eqref{x4}. Conical singularities can exist for arbitrary $\Lambda_1$. We note that generically the solution does not approach the minimum of $V_s$ close to the singularity. The generic solution close to a conical singularity has a nonvanishing $s''=U'=u$. The further evolution of $s$ towards the minimum is stopped by the vanishing of $\rho$, driven by the term $\sim U^{-1}$ in eq. \eqref{c5A}. For $\tilde \Lambda_1>3\tilde V/10$ the generic solution still has two singularities, but both can now be of the conical type. 

\begin{itemize}
\item [] {\bf b)\quad singularities with codimension larger than two}
\end{itemize}
We next turn to the case $D_2>1$ which corresponds to a codimension of the singularity larger than two. We also consider general $D_1$. For $\Lambda_2\neq 0$ modifications arise from the presence of the curvature of the $D_2$-dimensional subspace, $\Lambda_2$. We start with the direct product solution which exists for $\tilde V>0~,~\Lambda_1>0$, i.e. $\sigma(z)=\sigma_E$, as given by eq. \eqref{c12A}. We need to solve the remaining differential equation for $\rho$
\ba\label{c13}
\frac{\rho'}{\rho}&=&W=\pm \frac{2}{\sqrt{D_2(D_2-1)}}\sqrt{-F},\nn\\
F&=&\frac{2(D_2-1)}{D_1+D_2-1}\tilde V-\frac{D_2\Lambda_2}{\rho}.
\ea
For $\Lambda_2>0$ the range of $\rho$ is bounded, $0<\rho\leq\rho_{max}$,
\be\label{c14}
\rho_{max}=\frac{D_2(D_1+D_2-1)}{2(D_2-1)}\frac{\Lambda_2}{\tilde V},
\ee
with $W$ diverging at the boundary  $\rho\to 0$. 
One may be tempted to try the solution with a constant $\rho(z)=\rho_{max}$. This solves eq. \eqref{c13}, but for $U=W=0$ eq. \eqref{c7} is no longer linearly dependent and not obeyed for  $\tilde V\neq 0$. For $\rho=\rho_{\max}$ one has $W=0,\ W'=-2\Lambda_2/[(D_2-1)\rho_{max}]\neq 0$.

 The solution of eq. \eqref{c13} is symmetric under a reflection around the location of the maximum $z_{max},~\rho(z_{max})=\rho_{max}$. For $\rho\to 0$ the approximate equation $(\Lambda_2>0)$
\be\label{c15}
\rho'=\pm B\rho^{1/2}~,~B=\frac{2\sqrt{\Lambda_2}}{\sqrt{D_2-1}}
\ee
is solved by
\be\label{c16}
\rho=\frac{B^2}{4}z^2=\frac{\Lambda_2}{D_2-1}z^2
\ee
or by a similar behavior replacing $z\to(\bar z-z)$. For $\Lambda_2=D_2-1$ this is the regular behavior at the pole of the $D_2+1$-dimensional sphere. The direct product solutions are regular and there is no analogue to the conical singularity for $D_2=1$. With the replacement $\tilde V\to 4\tilde V/\big(D_2(D_1+D_2-1)\big)$ eq. \eqref{x8} is obeyed, but the integration constant is now fixed such that $\rho_0=1$ in eq. \eqref{x9}. For the unique direct product solution we therefore have $(\Lambda_2=D_2-1)$
\be\label{c16a}
\rho=\rho_{max}\sin^2(\sqrt{\rho_{max}}z).
\ee
This describes a sphere with radius $\sqrt{\rho_{max}}$.

For $D_2>1$ the direct product solution is regular. We may check that these regular solutions obey the brane constraint \eqref{B}. For the direct product solutions the coefficients $\tilde c_V,\ \tilde c_A$ are given by $\tilde c_V=\sigma_E\tilde V,\ \tilde c_A=\sigma_E\int_z\rho^{\frac{D_2}{2}-1}/\int_z\rho^{\frac{D_2}{2}}$ - a more general definition of these quantities will be given later. With $D=D_2+1,\ D_1=4,\ \Lambda_2=D_2-1,\ \Lambda_1=\Lambda$ the first equation in \eqref{B} corresponds to eq. \eqref{c12A}. The generalization of the second equation in \eqref{B}
\be\label{c16b}
\tilde c_A=\frac{2\tilde c_V}{(D_2-1)(D_1+D_2-1)}=\frac{D_2\sigma_E}{(D_2-1)\rho_{max}}
\ee
is indeed obeyed for the solution \eqref{c16a}.

Beyond the special direct product solutions we may consider eq. \eqref{c9} with arbitrary initial values for $s$ and $s'=U$. For a nontrivial warping we find generic solutions with singularities. 
The characteristic behavior close to a singularity where $\rho\to 0$ (or $\rho\to \infty$) is given by
\be\label{H1}
\rho=\rho_0z^\gamma~,~\sigma=\sigma_0z^{-\eta}~,~W=\frac{\gamma}{z}~,~U=-\frac{\eta}{z}.
\ee
For $z\to 0$, or similarly $(\bar z-z)\to 0$ if $z$ in eq. \eqref{H1} is replaced by $\bar z-z$, and $\gamma<2~,~\eta>-2$, the influence of $\Lambda_1,\Lambda_2$ and $\tilde V$ is subleading close to the singularity. The singular exponents $\gamma$ and $\eta$ are given by \cite{CWCC}
\ba\label{H2}
\eta_\pm &=&\frac{-2\pm 2\sqrt{D_2(D_1+D_2-1)/D_1}}{D_1+D_2},\nn\\
\gamma_\pm&=&\frac{1}{D_2}(2+D_1\eta_\pm).
\ea
The powerlike solutions \eqref{H1}, \eqref{H2} are exact solutions of the field equations for $\tilde V=\Lambda_1=\Lambda_2=0$. They have a brane or zerowarp singularity at $z=0$, dependending on the sign of $\eta_\pm$. The extension of this solution to $z\to\infty$ will be modified by nonzero $\tilde V,\Lambda_1$ or $\Lambda_2$. 

\begin{itemize}
\item []{\bf c)\quad solutions with two singularities}
\end{itemize}

For $\gamma<2$ a singularity encountered for decreasing $z$ (as in eq. \eqref{H1}) requires $\epsilon=+1$ in eq. \eqref{c5}, while a singularity for increasing $z$ needs $\epsilon=-1$. (This corresponds to antidamping at the singularity for the evolution of $s$ in eq. \eqref{c9}.) Indeed, using eq. \eqref{H2} we find for decreasing $z$
\be\label{H2a}
2-\gamma=\epsilon|\eta|\sqrt{D_1(D_1+D_2-1)/D_2}
\ee
and the opposite sign of $\epsilon$ for increasing $z$. Solutions with two singularities at $z=0$ and $z=\bar z$ therefore need a switch of $\epsilon$. If we define $\zeta={\rm sign}\ (\eta)$ we find that the singular exponents $(\eta_+,\gamma_+)$ correspond to $\epsilon\zeta=-1$ for $z$ increasing and to $\epsilon\zeta=1$ for $z$ decreasing. (Since $\eta_+>0$ one has $\zeta=1$.) The exponents  $(\eta_-,\gamma_-)$ correspond to the opposite  sign of $\epsilon\zeta$. (Now $\eta_-<0$ corresponds to $\zeta=-1$.) The exponents $(\eta_+,\gamma_+)$ correspond to a brane where $\rho\to 0,\ \sigma\to \infty$, while $(\eta_-,\gamma_-)$ describe a zerowarp with $\sigma\to 0,\ \rho\to \infty$.

For $\tilde V>0$ and $\Lambda_1>0$ we find
that the generic solutions interpolate between two singularities with opposite signs in eq. \eqref{H2}. They describe spaces with one brane at $z=0$ and one zerowarp at $z=\bar z$. This behavior is also found for $\Lambda_1=0$ or $\Lambda_1<0$. The brane singularity with exponents $(\eta_+,\gamma_+)$ can be replaced by a regular behavior $\rho\to z^2~,~\sigma\to\sigma_E$, if the integration constants are chosen appropriately. This special case corresponds to the geometries discussed in \cite{RDW}. For all solutions of this type or with two singularities we find a finite volume of internal space and a finite value of the four-dimensional gravitational constant, given by 
\be\label{H3}
c_R=\int^{\bar z}_0dz\rho^{\frac{D_2}{2}}\sigma^{\frac{D_1-2}{2}}. 
\ee
Geometries with two singularities are interesting because they lead to an effective $D_1$ dimensional gravity with finite $c_R$. 

We may ask what are the conditions for solutions with two singularities at $z=0$ and $z=\bar z$. (We include here the case where one singularity gets replaced by a regular behavior with $\rho\to 0$.) Solutions with two singularities need a switch of the sign $\epsilon$ between $\epsilon=1$ for $z\to 0$ and $\epsilon=-1$ for $z\to \bar z>0$. Continuity requires that such a switch can only occur if the square root in eq. \eqref{c5} vanishes at a point $z_c$ where
\be\label{49A}
U^2_c=\frac{4(D_2-1)F_c}{D_1(D_1+D_2-1)}.
\ee
Obviously, this is possible only if $F$ is positive. If the condition \eqref{49A} is met, not only $W$ but also all its derivatives remain continuous for solutions where $\epsilon$ switches sign. We can write eq. \eqref{c5} as 
\be\label{49B}
(D_2-1)W+D_1U=\epsilon\sqrt{K}/D_2,
\ee
such that the possible switch of $\epsilon$ for eq. \eqref{49A} occurs at $K=0$. A switch therefore always happens at an extremum of the combination $\rho^{D_2-1}\sigma^{D_1}$, and for solutions with two singularities this extremum has to be a maximum. Indeed, a switch from $\epsilon=1$ for $z<z_c$ to $\epsilon=-1$ for $z>z_c$ occurs if $(D_2-1)W'+D_1U'<0$ for $z=z_c$. With $\tilde V\geq 0$ and 
\ba\label{49C}
&&(D_2-1)W'+D_1U'=\frac{\epsilon K'}{2D_2\sqrt{K}}\nn\\
&&\qquad=\frac{2(D_2-1)\Lambda_2}{\rho_c}+
\frac{2D_1\Lambda_1}{\sigma_c}-4\tilde V\\
&&\qquad=-\frac{D_1(D_1+D_2-1)U^2_c}{2D_2}-\frac{4\tilde V}{D_2}
+\frac{2D_1\Lambda_1}{D_2\sigma_c}\nn
\ea
this quantity is negative. Indeed, the positivity of $F_c$ requires for $\Lambda_2>0$
\be\label{55A}
\sigma_c>\frac{D_1\Lambda_1}{2\tilde V}~,~\rho_c\geq
\frac{D_2\Lambda_2}{2\tilde V-D_1\Lambda_1/\sigma_c},
\ee
such that $(D_2-1)W'+D_1U'\leq-2\Lambda_2/\rho_c$.

For $\tilde V>0$ we have found that the solution with two singularities is indeed a generic behavior (for not too large $\Lambda_1$). For $\tilde V<0$, in contrast, the negative $d$-dimensional cosmological constant gives a negative contribution to $F$ in eq. \eqref{c4}. For positive $\Lambda_1$ and $\Lambda_2$ a switch of $\epsilon$ is no longer possible. In this case the generic solutions are expected to have only one singularity, while the volume of internal space and $c_R$ diverge for $z\to\infty$.  Such solutions do not lead to an effective $D_1$-dimensional gravity.

An interesting case is $\tilde V=0,\Lambda_1=0$. Then we find for $\Lambda_2>0$ no switch in $\epsilon$. The solutions correspond to a flat brane embedded in $d$-dimensional flat space. They are discussed in the appendix. In contrast, for $\Lambda_2<0$ one has positive $F$. Now we find again solutions with two singularities and a finite $c_R$. These solutions are also investigated in appendix A. Such solutions are interesting candidates for an effective $D_1$-dimensional gravity that has flat space as the static solution. The brane constraint \eqref{B}, i.e. $\Lambda_1\sim \tilde V$, will single out a vanishing $D_1$-dimensional cosmological constant $\Lambda_1=0$ if $\tilde V=0$. On the other hand, we will see that the second brane condition \eqref{B} can only be obtained for $\Lambda_2=0$ if $\tilde V=0$. We discuss the exact solutions for $\tilde V=0,\Lambda_1=0,\Lambda_2=0$ in the appendix. They are given for the whole range $0<z<\infty$ by eq. \eqref{H1}, with singular exponents \eqref{H2}. 

For a numerical investigation of solutions with two singularities one may take initial values at the switch point, with two free initial values $\sigma_{in}=\sigma_c~,~\rho_{in}=\rho_c$. The initial value $U_{in}=U_c$ is computed from eq. \eqref{49A}. The value $\rho_c$ has to obey eq. \eqref{55A}. For a fixed $\Lambda_2=D_2-1$ and positive $\Lambda_1\gtrsim 0$ we observe that small values $\tilde V\to 0$ imply large values of $\rho_c$. In the limit $\tilde V\to 0~,~\Lambda_1/\sigma_c\to 0$ the characteristic size of the $D_2$-dimensional subspace will diverge and the solution approaches the infinite volume solutions discussed in the appendix. A particular limiting case for the switch point is the direct product solution with
\ba\label{A9a}
\sigma_c&=&(D_1+D_2-1)\Lambda_1/(2\tilde V),\nn\\
\rho_c&\geq&
\frac{D_2(D_1+D_2-1)\Lambda_2}{2(D_2-1)\tilde V}=\rho_{max}.
\ea
Since $\rho\leq\rho_{max}$ is required by eq. \eqref{c13}, the switch point corresponds to $\rho_c=\rho_{max}$. 

The most general solution of the field equations has several integration constants. For a given $\Lambda_1$ we have four initial values $\sigma(z_0),~\sigma'(z_0),~\rho(z_0),~\rho'(z_0)$. Only three are independent due to eq. \eqref{c5} or \eqref{c5A}. One integration constant concerns only the overall normalization of $\sigma$. This has no physical relevance. For the spaces with finite volume we may fix the integration constant $\sigma(z_0)$ by requiring 
\be\label{44A}
\int dz\sigma^{\frac{D_1}{2}}\rho^{\frac{D_2}{2}}=1.
\ee
One further integration constant can be used to fix the location of one singularity (or of the regular point where $\rho=0$) at $z=0$. We therefore remain with one ``physical'' integration constant besides $\Lambda_1$. Typically, this determines the location of the (second) singularity at $\bar z$. 

\section{Models with scalar field}
\label{Models}
In this section we extend the discussion of singular solutions to higher dimensional gravity with a scalar field. For example, such a scalar field arises from a higher dimensional dilatation invariant theory \cite{9A}. In this case a nonvanishing scalar potential $\tilde V(\delta)$ will be produced only as a result of a dilatation anomaly. One purpose of this discussion is a demonstration that ``brane constraints'' of the type \eqref{B} arise generically for all higher dimensional fields. In general, the singular behavior is not directly linked to a conserved quantity. Avoiding the brane constraint  by a fixed brane tension is an option for gravity, where the singular behavior of the metric can be associated to a conserved energy momentum tensor. It seems much harder to defend a similar option for a scalar field since no associated conserved charge is available. 

As a second purpose, we investigate possible modifications of the singular behavior \eqref{H1}, \eqref{H2}. Indeed, in presence of a scalar field there are new possibilities for an asymptotic behavior of the solutions for $z\to\infty$ or for $z\to 0$ or $z\to \bar z$. As an example, we  concentrate on an exponential potential
\be\label{s1}
\tilde V(\delta)=\tilde v_0\exp(-\alpha\delta).
\ee
It admits scaling solutions of the type
\ba\label{s2}
\tilde V&=&\tilde v_sz^{-2}~,~\delta=\delta_0+\frac2\alpha \ln z~,~\delta_0=\frac1\alpha 
\ln \frac{\tilde v_0}{\tilde v_s},\nn\\
\delta'^2&=&\frac{4}{\alpha^2}z^{-2}~,~\delta''=-\frac{2}{\alpha}z^{-2}~,~\nn\\
\sigma&=&\sigma_0 z^{-\eta}~,~\rho=\rho_0z^\gamma.
\ea
(Here $z$ may be replaced by $\bar z-z$.) The scalar field equation is obeyed for 
\be\label{s3}
\tilde v_s=\frac{1}{\alpha^2}(2+D_1\eta-D_2\gamma)=
\frac{2\kappa}{\alpha^2}.
\ee
For $\tilde v_0>0$ such solutions can only exist for
\be\label{s4}
\kappa=1+\frac{D_1\eta}{2}-\frac{D_2\gamma}{2}>0.
\ee
whereas negative $\tilde v_0$ implies negative $\kappa$. 
The corresponding behavior of the volume element 
\be\label{s5}
dz\rho^{D_2/2}\sigma^{D_1/2}\sim dz z^{1-\kappa}
\ee
leads to a finite volume for $\kappa<2$ for $z\to 0$ (or $z\to\bar z$) or for $\kappa>2$ for $z\to\infty$. For a finite gravitational constant after dimensional reduction the conditions are $\kappa<2+\eta~ (z\to0)$ or $\kappa>2+\eta~(z\to\infty)$. 

We want to relate the exponents $\eta,\gamma$ and $\kappa$ to the behavior of the potential characterized by $\alpha$. For this purpose we first investigate the relations between $\eta,\gamma$ and $\tilde v_s$. 
For $\sigma$ and $\rho$ decreasing sufficiently slowly or fast, the terms $\sim\Lambda_{1,2}$ can be neglected in eqs. \eqref{c6}, \eqref{c7},which yield the relations
\ba\label{s9}
\eta&+&\frac{D_1}{2}\eta^2-\frac{D_2}{2}\eta\gamma+\frac{4}{D_1+D_2-1}\tilde v_s=0,\\
-\gamma&+&\frac{D_2}{2}\gamma^2 -\frac{D_1}{2}\eta\gamma+\frac{4}{D_1+D_2-1}\tilde v_s=0,
\ea
and therefore
\be\label{s11}
\eta+\gamma+\frac12(D_1\eta^2-D_2\gamma^2)+\frac12(D_1-D_2)\eta\gamma=0.
\ee
Eq. \eqref{s9} generalizes eq. \eqref{H2}
\be\label{s12}
D_2\gamma=2+D_1\eta+\frac{8}{D_1+D_2-1}\frac{\tilde v_s}{\eta},
\ee
and implies
\be\label{s13}
\kappa=1+\frac{D_1}{2}\eta-\frac{D_2\gamma}{2}=-\frac{4}{D_1+D_2-1}
\frac{\tilde v_s}{\eta}.
\ee
Since $\kappa$ and $\tilde v_s$ have the same sign, cf. \eqref{s3}, this type of scaling behavior is possible only for $\eta<0$. The singularities are zerowarps as far as the behavior of $\sigma$ is concerned. 

Combining eqs. \eqref{s13}, \eqref{s3} yields the relation between $\eta$ and $\alpha$
\be\label{s14}
\eta=-\frac{8}{(D_1+D_2-1)\alpha^2}.
\ee
The relation between $\gamma$ and $\eta$ can then be inferred from eq. \eqref{s11}. This quadratic equation has two solutions
\be\label{s15}
\gamma_1=\frac{1}{D_2}(2+D_1\eta)~,~\gamma_2=-\eta.
\ee
The first solution $\gamma_1$ corresponds to $v_s=0$. In this case $\tilde V$ decays faster then $\sim z^{-2}$ and plays no role asymptotically, such that eq. \eqref{H2} is recovered. The new type of scaling solution, for which the potential matters, obeys
\be\label{s16}
\gamma=-\eta=\frac{8}{(D_1+D_2-1)\alpha^2},
\ee
and therefore
\ba\label{s17}
\kappa&=&1-\frac{D_1+D_2}{2}\gamma=1-\frac{4(D_1+D_2)}{(D_1+D_2-1)\alpha^2},\nn\\
\kappa-\eta&=&1-\frac{4(D_1+D_2-2)}{(D_1+D_2-1)\alpha^2}.
\ea
For all $\alpha$ one finds $\kappa<1~,~\kappa-\eta<1$. A singularity a $z\to 0$ with
\be\label{s18}
\sigma=\bar\sigma z^\gamma~,~\rho=\bar\rho z^\gamma~,~\gamma>0,
\ee
leads to a finite volume and a finite gravitational constant of the effective $D_1$-dimensional theory. Since $\rho$ vanishes for $z\to 0$ this is a brane singularity combined with a zerowarp.  We infer that the singular exponents near a brane singularity can be modified by the presence of a scalar field.

The scaling behavior \eqref{s16}, \eqref{s18} implies $U,W\to \infty$ as the singularity is approached. It corresponds to the plus sign for the square root in eq. \eqref{s7}, and to the plus sign in eq. \eqref{c5} $(\epsilon=+1)$. A similar singularity can be reached for $z\to \bar z$. (Now $\epsilon=-1$ in eq. \eqref{c5}.)  A numerical solution of the system of differential equations finds indeed solutions with two such singularities if $\alpha^2>4/(D_1+D_2-1)$. For generic values of the integration constants these seem to be the characteristic solutions for $\Lambda_1=0$, but also for positive and negative $\Lambda_1$. Such geometries with two cups seem perfectly acceptable. We note that for both singularities $\delta$ diverges logarithmically to negative values, such that $V(\delta)$ diverges. 

One may also check the consistency of the singular behavior \eqref{s16} with eq. \eqref{c5} 

If $\Lambda_1$ and $\Lambda_2$ can be neglected, one finds the asymptotic behavior 
\be\label{s6}
F=\left(2\tilde v_s-\frac{4}{\alpha^2}\right)z^{-2}=
\frac{2}{\alpha^2}(D_1\eta-D_2\gamma)z^{-2}.
\ee
For $D_2>1$ eq. \eqref{c5} relates $\gamma$ to $\eta$
\ba\label{s7}
&&(D_2-1)\gamma=D_1\eta\pm\frac{1}{\sqrt{D_2}}\\
&&\sqrt{D_1(D_1+D_2-1)\eta^2-8(D_2-1)(D_1\eta-D_2\gamma)/\alpha^2}.\nn
\ea
The argument of the square root must be positive, requiring
\be\label{s8}
\alpha^2\geq
\frac{8(D_2-1)(D_1\eta-D_2\gamma)}{D_1(D_1+D_2-1)\eta^2}.
\ee

The condition for the neglection of the terms $\sim\Lambda_2\rho^{-1}$ near the singularity requires $\gamma<2$ or $\alpha>\alpha_c$, 
\be\label{s19}
\alpha_c=\frac{2}{\sqrt{D_1+D_2-1}}.
\ee
For $\alpha<\alpha_c$ the behavior at the singularities is different. For example, a numerical integration of the field equations shows solutions where $\eta$ vanishes at two zerowarp singularities at $z=0,~z=\bar z$, while $\rho$ diverges. 

\section{Dimensional reduction}
\label{Dimesionalreduction}
\begin{itemize}
\item []{\bf a)\quad higher dimensional gravity}
\end{itemize}

If internal space is in some sense small as compared to a typical wavelength of an experiment, say at the LHC, we do not expect that we can resolve it. In this case one expects a valid description in terms of an effective four-dimensional theory, obtained by taking a suitable average over the hidden internal dimensions. In fact, a higher dimensional theory can always be formulated as a four-dimensional theory with infinitely many degrees of freedom. The effective four-dimensional theory obtains if one only keeps a finite number of these modes. Usually, the neglection of the infinitely many other modes is related to the heavy mass of these ``Kaluza-Klein particles'' - often many orders of magnitude larger than the inverse of the experimental wave length. For a given model and solution, this neglection has, of course, to be justified.

Dimensional reduction to a $D_1$-dimensional theory obtains by expanding the higher dimensional field in some complete set of functions depending on the internal coordinates $z$ and $y^\alpha$. The coefficients depend on the $D_1$-dimensional coordinates $x^\mu$ and correspond to the $D_1$-dimensional fields. There is one field for each basis function in the complete system of functions. The effective $D_1$-dimensional action obtains by integrating the $d$-dimensional action over the internal coordinates $z$ and $y^\alpha$. At this stage it is a functional of infinitely many $D_1$-dimensional fields. It is completely equivalent to the higher dimensional action - we have just chosen a suitable representation of the $d$-dimensional functions. In particular, there is a one to one correspondence between an extremum of the $d$-dimensional action and an extremum of the $D_1$-dimensional action with respect to variations of all the infinitely many fields. Of course, if there are constraints limiting the allowed variations of the $d$-dimensional fields - as for fixed tension branes - these constraints have also to be implemented on the level of the $D_1$-dimensional fields.

Let us consider variable tension branes with unconstrained $d$-dimensional metric and scalar field. The extremum condition for the action implies an extremum with respect to the variation of any {\em single} $D_1$-dimensional field. We will truncate the $D_1$-dimensional action by keeping only a finite number of fields that correspond to fluctuations around some given solution to the $d$-dimensional field equation. If this $d$-dimensional solution corresponds to an extremum of the $d$-dimensional action, the truncated effective $D_1$-dimensional action must also have an extremum for those values of the $D_1$-dimensional fields that correspond to a vanishing fluctuation around the solution. We will find that this is not the case for arbitrary integration constants of the solutions discussed in sects. \ref{Einsteinspaces} and \ref{Models}. There is a clash between the $d$-dimensional solutions and the $D_1$-dimensional solutions.

We begin by considering only fluctuations of the $D_1$-dimensional metric. The dimensionally reduced $D_1$-dimensional action for the metric $g^{(D_1)}_{\mu\nu}(x)$ obtains by inserting a solution for $\sigma(z),\rho(z),\delta(z)$ and integrating over the internal $D_2+1$ coordinates $y,z$. It has the form
\be\label{T1}
S{[g^{(D_1)}\mn]}=\int_x(g^{(D_1)})^{1/2}
\left\{\frac{\M}{2}R^{(D_1)}-V^{(D_1)}\right\},
\ee
with effective reduced $D_1$-dimensional Planck mass $M$,
\be\label{T2}
\M=c_R\Omega_{D_2}
M^{d-2}_d~,~
c_R=\int_z\rho^{\frac{D_2}{2}}
\sigma^{\frac{D_1-2}{2}},
\ee
and $\Omega_{D_2}$ is the volume of the $D_2$-dimensional subspace with metic $g^{(D_2)}_{\alpha\beta}$. The effective potential or $D_1$-dimensional cosmological constant $V\de$ reads
\be\label{T3}
V\de=\Omega_{D_2}
M^{d-2}_d\int_z
\rho^{\frac{D_2}{2}}
\sigma^{\frac{D_1}{2}}
\left\{\tilde V+\frac12\delta'^2-\frac12\tilde R_{int}\right\}.
\ee
Here $\tilde R_{int}$ is the ``internal space contribution'' to the higher dimensional curvature scalar
\be\label{T4}
\tilde R_{int}=\tilde R_{sol}-D_1\Lambda_1\sigma^{-1},
\ee
with $\tilde R_{sol}$ the curvature scalar for the solution around which we expand. 
(While we keep $\rho,\sigma$ and $\delta$ fixed, we consider the ansatz \eqref{A1} with arbitrary $g\de\mn (x)$, such that $\tilde R=\tilde R_{int}+R\de\sigma^{-1}$.) The values for $M$ and $V^{(D_1)}$ will depend on the particular solution of the higher dimensional field equations $(\rho,\sigma,\delta)$ for which we perform the dimensional reduction. 

The internal curvature contribution to $V^{(D_1)}$ involves the $D_2+1$ dimensional curvature scalar
\be\label{T5}
R^{(D_2+1)}=D_2
\left\{\frac{\Lambda_2}{\rho}-\frac{\rho''}{\rho}-\frac14(D_2-3)
\left(\frac{\rho'}{\rho}\right)^2\right\}
\ee
and a contribution from the warping
\ba\label{T6}
&&\tilde R_{int}=R^{(D_2+1)}\nn\\
&&-D_1\left\{\frac{\sigma''}{\sigma}+\frac14(D_1-3)\left(\frac{\sigma'}{\sigma}\right)^2+\frac12D_2
\frac{\rho'}{\rho}\frac{\sigma'}{\sigma}\right\}.
\ea
For all solutions to the higher dimensional field equations it obeys
\be\label{T7}
\tilde R_{int}=\delta'^2+\frac{2(D_1+D_2+1)}{D_1+D_2-1}\tilde V-D_1\frac{\Lambda_1}{\sigma},
\ee
such that
\ba\label{T8}
V\de&=&\Omega_{D_2}M^{d-2}_d\int_z\rho^{\frac{D_2}{2}}\sigma^{\frac{D_1}{2}}
\left\{\frac{D_1\Lambda_1}{2\sigma}-\frac{2\tilde V}{D_1+D_2-1}\right\}\nn\\
&=&\frac{D_1\Lambda_1\M}{2}-\frac{2\M}{D_1+D_2-1}\frac{c_V}{c_R}~,
\ea
with
\be\label{71A}
c_V=\int_z\rho^{\frac{D_2}{2}}
\sigma^{\frac{D_1}{2}}\tilde V.
\ee
If we insert in eq. \eqref{T1} the metric $g^{(D_1)}\mn$ corresponding to the solutions of the higher dimensional field equations, i.e. $R^{(D_1)}=D_1\Lambda_1$, we find with eqs. \eqref{T2}, \eqref{T8} the result \eqref{c04}, as it should be. 

From eq. \eqref{T1} we can compute the field equations for the metric $g^{(D_1)}\mn$ in the dimensionally reduced theory
\be\label{T9} 
\M(R^{(D_1)}\mn-\frac12 R^{(D_1)}g^{(D_1)}\mn)=-V^{(D_1)} g^{(D_1)}\mn,
\ee
which yields for the curvature scalar
\be\label{T10}
R^{(D_1)}=\frac{2D_1}{D_1-2}\frac{V^{(D_1)}}{\M}.
\ee
In general, this does not have the solution $R\de=D_1\Lambda_1$, as required by the $d$-dimensional solution. We observe a clash between the extremum of the $D_1$-dimensional action, given by eq.\eqref{T10}, and the solution of the $d$-dimensional field equations. Only for special values of the integration constants of the higher dimensional solution the two will coincide.

Consistency of the higher dimensional solution $(R^{(D_1)}=D_1\Lambda_1)$ with the effective $D_1$-dimensional theory requires
\be\label{T11}
\frac{V^{(D_1)}}{\M}=\frac{D_1-2}{2}\Lambda_1.
\ee
Using eqs. \eqref{T2}, \eqref{T8} one finds
\be\label{T12}
\frac{V^{(D_1)}}{\M}=\frac{D_1\Lambda_1}{2}-\frac{2c_V}{(D_1+D_2-1)c_R}
\ee
and therefore the constraint
\be\label{T13}
\Lambda_1=\frac{2c_V}{(D_1+D_2-1)c_R}. 
\ee
We recognize the brane constraint \eqref{B}, with $\Lambda_1=\Lambda,D_1=4,\tilde c_V=c_V/c_R$. If this constraint is not obeyed, the higher dimensional action is not an extremum with respect to local variations of the $D_1$-dimensional metric $g^{(D_1)}\mn$. This variation is performed with all other metric components and the scalar field, i.e. the functions $\rho,\sigma,\delta$, kept fixed at the values corresponding to a solution of the higher dimensional field equations. One concludes that the solution to the higher dimensional field equation is a necessary, but not a sufficient condition for an extremum of $S$. An extremum with respect to arbitrary variations requires in addition the brane constraint \eqref{T13}. 

Similar considerations can be made for other variations. For example, we may vary the volume of internal space by multiplying $\tilde g_{\alpha\beta}$ and $\tilde g_{zz}$ by a common factor $L^2$. We want to compute the dependence of the action on $L$, while keeping $\rho(z),\sigma(z),\delta(z)$ as well as $g^{(D_1)}\mn$ and $g^{(D_2)}_{\alpha\beta}$ at fixed values. The volume factor yields a factor $L^{D_2+1}$, while the different contributions to $S$ scale differently: $\partial^{\hat\rho}\delta\partial_{\hat\rho}\delta\sim L^{-2}~,~\tilde R_{int}\sim L^{-2}~,~R^{(D_1)}\sigma^{-1}\sim L^0~,~\tilde V\sim L^0$. As a result, the dependence of $S$ on $L$ is not trivial
\ba\label{T14}
S(L)&=&\int_x(g^{(D_1)})^{1/2}\M L^{D_2+1}
\left\{\frac{D_1\Lambda_1}{2}(1-L^{-2})\right.\nn\\
&&\left.-\frac{c_V}{c_R}+\frac{D_1+D_2+1}{D_1+D_2-1}
\frac{c_V}{c_R}L^{-2}\right\}.
\ea
The solution of the $d$-dimensional field equations corresponds to $L=1$. Again, the extremum of $S(L)$ does generically not occur for $L_0=1$, and a constraint on the integration constants of the solution is needed in order to avoid a clash between the extremum of $S(L)$ and the $d$-dimensional solution. 

An extremum with respect to $L$ requires for $L_0=1$
\ba\label{T15}
&&\hspace{0.3cm} (D_2+1)
\left(\frac{D_1\Lambda_1}{2}-\frac{c_V}{c_R}\right)\nn\\
&&-(D_2-1)
\left(\frac{D_1\Lambda_1}{2}-\frac{D_1+D_2+1}{D_1+D_2-1}
\frac{c_V}{c_R}\right)=0
\ea
or
\be\label{T16}
\Lambda_1=\frac{2c_V}{(D_1+D_2-1)c_R}. 
\ee
This is the same condition as \eqref{T13}. Thus, if the brane constraint \eqref{T13} is violated, the action is not an extremum with respect to variations of the volume of internal space either. We could also consider a class of metrics where the overall scale of $\sigma$ is varied, $\sigma(z)\to A_\sigma\sigma(z)$. With 
\ba\label{T17}
S(A_\sigma)&=&\int_x(g^{(D_1)})^{1/2}
\M A^{\frac{D_1}{2}}_\sigma
\left\{\frac{D_1\Lambda_1}{2}(A^{-1}_\sigma-1)\right.\nn\\
&&\left.+\frac{2c_V}{(D_1+D_2-1)c_R}
\right\}
\ea
the condition $\partial S(A_\sigma)/\partial A_{\sigma|A_\sigma=1}=0$ yields again the brane constraint \eqref{T13}. 

Similarly, we may vary the overall scale of $\rho$, i.e. $\rho(z)\to A_\rho\rho(z)$, while keeping all other functions fixed. From 
\ba\label{T23}
&&S[A_\rho]=\M\int_x(\tilde g^{(D_1)})^{1/2}
A_\rho^{\frac{D_2}{2}}\nn\\
&&\left\{\frac{D_2\Lambda_2}{2}\left(\frac{1}{A_\rho}-1\right)
\frac{c_\rho}{c_R}+\frac{2c_V}{(D_1+D_2-1)c_R}
\right\},
\ea
with
\be\label{74A}
c_\rho=\int_z\rho^{\frac{D_2-2}{2}}\sigma^{\frac{D_1}{2}},
\ee
we obtain the condition for an extremum at $A_\rho=1$,
\be\label{T25}
\Lambda_2c_\rho=\frac{2}{D_1+D_2-1}c_V.
\ee
This corresponds to the second brane condition in eq. \eqref{B}, with $\tilde c_A=c_\rho/c_R$. The combination with eq. \eqref{T13}, which corresponds to a volume conserving opposite multiplicative scaling of $\sigma$ and $\rho$, yields 
\be\label{T26}
\Lambda_2c_\rho=\Lambda_1c_R. 
\ee
Eq. \eqref{T25} has an important consequence for $D_2=1$ where $\Lambda_2=0$. Now $S[A_\rho]$ is a monotonic function of $A_\rho$ for all $\tilde V\neq 0$. As a consequence, the extremum condition \eqref{T25} can be obeyed only for $\tilde V=0$. We will understand better in sects. \ref{Cosmologywith}, \ref{runawaysolutions} why this excludes all singular finite volume solutions if metric variations corresponding to a constant multiplicative rescaling of $\rho$ are allowed. 

We conclude that for geometries with singularities there are additional conditions for the extrema of the action, as eq. \eqref{T13}, \eqref{T25}, that go beyond the solution of the field equations. They are associated to ``global variations'' of the fields, while the field equations are the extremum condition only for local field variations. In particular, for the class of models considered here the consistent extrema with flat $D_1$-dimensional space, $\Lambda_1=0$, require $\tilde V=0$ and, by eq. \eqref{T26}, also $\Lambda_2=0$. These solutions are given by flat space or the exact singular solutions \eqref{H1}, \eqref{H2}. 

\begin{itemize}
\item []{\bf b)\quad higher dimensional scalar fields}
\end{itemize}

We may also investigate global variations of the scalar field $\delta$. One possibility is an overall constant shift $\delta\to\delta+\Delta$. This affects only the potential term. The condition that a given solution of the higher dimensional field equations is also an extremum with respect to the global shifts is simply
\be\label{T18}
\int_z\rho^{\frac{D_2}{2}}\sigma^{\frac{D_1}{2}}
\frac{\partial\tilde V}{\partial\delta}=0. 
\ee
For the exponential potential \eqref{s1} it can only be obeyed for $\delta\to \infty$ where $\tilde V(\delta)\to 0$.

Another possibility are multiplicative rescalings $\delta\to A_\delta \delta$, with 
\ba\label{T19}
S(A_\delta)&=&\M\int_x(g^{(D_1)})^{1/2}
\left\{(1-A^2_\delta)\frac{c_K}{c_R}\right.\\
&&\left.+\frac{D_1+D_2+1}{D_1+D_2-1}\frac{c_V}{c_R}
-\frac{1}{c_R}\int_z\rho^{\frac{D_2}{2}}
\sigma^{\frac{D_1}{2}}\tilde V(A_\delta\delta)\right\},\nn
\ea
where
\be\label{T20}
c_K=\frac12\int_z\rho^{\frac{D_2}{2}}\sigma^{\frac{D_1}{2}}\delta'^2.
\ee
The extremum condition $\partial S/\partial A_\delta|_{A_\delta=1}$ reads
\be\label{T21}
\int_z\rho^{\frac{D_2}{2}}\sigma^{\frac{D_1}{2}}
\frac{\partial\tilde V}{\partial\delta}\delta=-2c_K.
\ee
Using the scalar field equation \eqref{c02} and the definition of the covariant Laplacian $\hat D^2$ we can write the condition \eqref{T21} as 
\be\label{T22}
\int_z\frac{\partial}{\partial z}
\left\{\rho^{\frac{D_2}{2}}\sigma^{\frac{D_1}{2}}\delta\delta'\right\}=0.
\ee
It is obviously obeyed for compact spaces, whereas in presence of singularities it contains possible boundary terms.

We can construct the effective $D_1$-dimensional action for a scalar field associated to local multiplicative variations of $\delta$ by using the ansatz
\be\label{95Aa}
\delta(z,y,x)=\delta(z)\varphi_\delta(x).
\ee
The solution to the higher dimensional field equations corresponds to a homogeneous and static field $\varphi_\delta(x)=1$. On the other hand, one finds for the $D_1$-dimensional action (with fixed metric $g\de\mn$ corresponding to the higher dimensional solution)
\be\label{95B}
S[\vd]=-\int_x(g\de)^{1/2}\Big\{V_\delta(\vd)
+\frac12c_\delta\partial^\mu\vd\partial_\mu\vd\Big\}
\ee
with
\ba\label{95C}
V_\delta(\vd)&=&\int_z\sigma^{\frac{D_1}{2}}\rho^{\frac{D_2}{2}}
\Big\{\tilde V(\delta\vd)+\frac12\vd^2\delta'^2\Big\},\nn\\
c_\delta&=&\int_z\sigma^{\frac{D_1-2}{2}}\rho^{\frac{D_2}{2}}\delta^2.
\ea
The $D_1$-dimensional field equations read
\ba\label{95D}
&&D^\mu D_\mu\vd=c^{-1}_\delta\frac{\partial V_\delta}{\partial\vd}=c^{-1}_\delta\int_z\left[\partial_z\left(\sigma^{\frac{D_1}{2}}\rho^{\frac{D_2}{2}}\delta'\delta\right)\right.\nn\\
&&\left.+\sigma^{\frac{D_1}{2}}\rho^{\frac{D_2}{2}}
\left(\delta'^2+\frac{\partial^2\tilde V}{\partial \delta^2}\right)(\varphi_\delta-1)
+\dots\right].
\ea
Here the dots denote higher orders in an expansion in $\varphi_\delta-1$. For a static solution at $\varphi_\delta=1$ the first term in eq. \eqref{95D} has to vanish, cf. eq. \eqref{T22}. The second term $\sim(\varphi_\delta-1)$ determines the mass of the scalar field $\varphi_\delta$.

\section{Field variations}
\label{fieldvariations}
\begin{itemize}
\item []{\bf a)\quad brane constraints and field variations}
\end{itemize}

One may ask if the variations of the metric and the scalar field considered in the preceding section are legitimate. Consider a higher dimensional solution $(\rho,\sigma,\delta)$ for which $c_R$ and $c_V$ are finite. It is then easy to see that there exist variations of the $d$-dimensional metric $\tilde g_{\hat\mu\hat\nu}$ for which the action remains finite, while the solution of the field equations is not an extremum with respect to such variations. One example are variations of the metric $g^{(D_1)}_{\mu\nu}$ which are local in the $D_1$-dimensional space. They result in a finite change of the dimensionally reduced action \eqref{T1} and therefore also in a finite change of the higher dimensional action. As another example, we may replace the global change of the ``size of internal space'', $L$, by a change $L(x)$ which is local in the $D_1$ coordinates $x^\mu$. This results in replacing eq. \eqref{T14} by 
\ba\label{T27}
S&=&-\int_x(g^{(D_1)})^{1/2}
\big\{V^{(D_1)} (L)\nn\\
&&+c_L\M L^{D_2+1}
\partial^\mu\ln L~\partial_\mu \ln L\big\}.
\ea
We can construct local variations of $L(x)$ for which the ``kinetic term'' $\partial^\mu\ln L\partial_\mu \ln L$ is negligible, while the action remains finite. The extremum condition for such variations is then the condition that the potential $V\de(L)$ should have an extremum for the value $L=L_0=1$, around which the variation is taken. This condition, $\partial V^{(D_1)} /\partial L|_{L_0=1}=0$, is precisely the brane constraint \eqref{T13}. If we look for an extremum with respect to all allowed variations of the higher dimensional metric, which change the action by a finite amount, we have to impose the brane constraint, in addition to the local field equations.

One should interprete the higher dimensional action as the quantum effective action after including all quantum fluctuations. (Formally, it is the generating functional of the ``full'' one particle irreducible vertices.) Physical states correspond to an extremum of the quantum effective action with respect to all allowed variations of $\tilde g_{\hat\mu\hat\nu}$ and $\delta$ for which it is well defined. This is the case for all variations which change the action only by a finite amount. It is therefore difficult to argue why the condition \eqref{T13} should be dismissed unless the model contains some additional ingredients which restrict the allowed values of the metric and scalar field, as for fixed tension branes. 

For the higher dimensional Einstein spaces discussed in sect. \ref{Einsteinspaces} one finds finite $c_R$ and $c_V$. The brane constraint \eqref{T13} selects among the solutions of the higher dimensional field equations the ones with $\Lambda_1$ given by $c_V$ and $c_R$. We recall that $c_R$ is always positive and $c_V$ vanishes only for $\tilde V=0$. For a nonvanishing higher dimensional cosmological constant $\tilde V\neq 0$ one therefore selects solutions with $\Lambda_1\neq 0$. Furthermore, also $c_\rho$ \eqref{74A} is finite for the solutions with finite volume discussed in sect. \ref{Einsteinspaces}. If we also impose eq. \eqref{T25} (with a fixed value of $\Lambda_2$, say $\Lambda_2=D_2-1$) this yields an additional relation between $c_V$ and $c_\rho$ and therefore, for given $\tilde V$, a further brane constraint on the functions $\sigma(z)$ and $\rho(z)$.

This implies that a continuous family of solutions with arbitrary $\Lambda_1$ remains no longer possible. Indeed, for a given value of $\Lambda_1$ two integration constants of the most general solution are fixed by the brane constraints \eqref{T13} and \eqref{T25}. As mentioned before, the overall scale of $\sigma$ is arbitrary and we may fix it by the condition $c_V=\tilde V$. One further integration constant can be used to have the location of one singularity at $z=0$. We then have only one independent integration constant left, but two constraints. As a result, for an arbitrary value of $\Lambda_1$ the system is overdetermined. We expect that solutions exist at most for discrete values of $\Lambda_1$, for which the $\Lambda_1$-dependent integral $c_R(\Lambda_1)$ matches eq. \eqref{T13}. Indeed, the relation \eqref{T25} fixes all integration constants for given $\Lambda_1$, such that $c_R(\Lambda_1)$ is a given function. The discrete values for $\Lambda_1$ must then obey $(D_1+D_2-1)c_R(\Lambda_1)\Lambda_1=2\tilde V$. It is not clear, a priori, if such solutions with finite volume exist - this depends on the range that $c_\rho(\Lambda_1)$ can take for a given $\tilde V$. If not, only the infinite volume spaces, as the $D_1+D_2+1$-dimensional de Sitter space, are extrema with respect to arbitrary variations of the metric. 

The condition for a finite volume solution with two singularities can be written as
\be\label{92A}
\frac{\int_z\rho^{\frac{D_2-2}{2}}\sigma^{\frac{D_1}{2}}}{\int_z\rho^{\frac{D_2}{2}}\sigma^{\frac{D_1}{2}}}
=\frac{2\tilde V}{(D_1+D_2-1)\Lambda_2},
\ee
For $D_2>1$ and $\Lambda_2\neq 0$ the overall scale of $\rho$ is $\sim\Lambda_1^{-1}$ according to eq. \eqref{T26}. If the dimensionless function $c_R(\Lambda_1)$ does not vary too strongly (for given $\tilde V$), such that $c_R(\Lambda_1)\Lambda_1$ remains a monotonic function of $\Lambda_1$ which vanishes for $\Lambda_1\to 0$ and diverges for $\Lambda_1\to\infty$, one will find exactly one solution of eq. \eqref{T13}. We recall that eq. \eqref{T25} has no solution for $D_2=1~(\Lambda_2=0)$ if $\tilde V\neq 0$. Our discussion of spaces with singularities includes as special cases the regular behavior at $z=0$ and $z=\bar z$. We have found in sect. \eqref{Einsteinspaces} special values for the integration constants where the geometry is a direct product of $D_1$-dimensional space and a compact (regular) internal space. This solution obeys the brane constraint. It is tempting to speculate that these regular solutions are the only extrema of the higher dimensional action with finite $c_V$ and $c_R$. 

The situation is more complicated for the solutions with a scalar field discussed in sect. \ref{Models}. For $\tilde V>0$ and for solutions showing at the singularities the scaling behavior \eqref{s2}, the integral $c_V$ is not a finite quantity anymore. Near a singularity $\tilde V$ diverges $\sim z^{-2}$, or 
\be\label{91}
c_V\sim\int dz z^{-(1+\kappa)},
\ee
and $\kappa>0$ if $\tilde v_0>0$. As a consequence, the effective $D_1$-dimensional theory does not exist - the potential $V$ in eq. \eqref{T8} is not well defined. The arguments leading to the constraint \eqref{T13} are not directly applicable in this case. This remark also holds for the constraint \eqref{T25}. With $c_\rho\sim \int_zz^{1-\kappa-\gamma}\sim\int_z z^{(D_1+D_2-2)\gamma/2}$ finite, we can still perform the variations which require the constraint \eqref{T26} as an extremum condition. On the other hand, for $\tilde V<0$ one has $\kappa<0$ and $c_V$ is well defined. The brane constraints \eqref{T13}, \eqref{T25} are now well defined  and have to be obeyed for variable tension branes. Furthermore, for an exponential potential the constraint \eqref{T18} reads $\alpha c_V=0$ and is well defined. Its only solution is $c_V=0$ or $\delta_0\to \infty,\tilde v_s\to 0$. We note that for $\kappa<1$ the condition \eqref{T22} is obeyed for all singular solutions of the type \eqref{s2}. We conclude that the issue of the scalar brane constraints \eqref{T18} and \eqref{T22} depends on the detailed shape of the potential and solution. We will not discuss further here the issue how to deal with the case where $c_V$ is not finite.

We may discuss the variations that are not accounted for by the field equations in a more general context. This is most easily done for the scalar fields. With respect to some general variation $\delta\to\delta+\epsilon$ the variation of the $d$-dimensional action is given by eq. \eqref{92}. Since the field equations are obeyed, the additional constraints arising for non-local variations $\epsilon$ read
\be\label{95AA}
\int\partial_{\hat\rho}\{\tilde g^{1/2}\epsilon\partial^{\hat\rho}\delta\}=0.
\ee
For example, taking constant $\epsilon$ leads to the constraint \eqref{T18} 
\be\label{93}
\int_z\sigma^{D_1/2}\rho^2\hat D^2\delta=\int_z\sigma^{D_1/2}\rho^{D_2/2}
\frac{\partial\tilde V}{\partial \delta}=0.
\ee
Similarly, for $\epsilon=A\delta$, with constant $A$, the extremum condition for the action yields the constraint \eqref{T22}.

The situation with respect to variations of the metric $\tilde g_{\hat\mu\hat\nu}$ is analogous. Again, the field equations \eqref{c01} guarantee an extremum of the action only in the absence of boundary contributions. In the presence of singularities one may require that the action is an extremum also with respect to metric variations which do not lead to vanishing boundary terms. This leads to the brane constraints \eqref{T13}, \eqref{T25}, \eqref{T26}. For example, a variation of the $D_1$-dimensional metric $g^{(D_1)}_{\mu\nu}(x)$, even if local in the $D_1$ coordinates $x^\mu$, amounts still to a universal variation for all coordinates of the $D$-dimensional internal space. This variation does not vanish at the location of the singularities and gives a nonvanishing contribution to the boundary term. We conclude that the additional constraints beyond the field equations are connected to the extremum condition for the metric with respect to variations that are not local in the higher dimensional space time and do not vanish at the singularities. 

We may explicitely compute $K^{\hat\nu}$ in eq. \eqref{KXA} for appropriate metric variations. For example, taking $\tilde h\ab=(A_\rho-1)\rho(z)g\dz\ab(y)$ (and vanishing other components), one finds the non-vanishing component
\be\label{G14a}
K^z=\frac{D_2}{2}\frac{\rho'}{\rho}(A_\rho-1)
\ee
and therefore the extremum condition
\ba\label{G14b}
&&\int_z\partial_z\left(\sigma^{\frac{D_1}{2}}\rho^{\frac{D_2-2}{2}}\rho'\right)
=\int_z\sigma^{\frac{D_1}{2}}\rho^{\frac{D_2}{2}}\\
&&\qquad\left(\frac{D_1}{2}\frac{\sigma'}{\sigma}\frac{\rho'}{\rho}+\frac{D_2-2}{2}
\left(\frac{\rho'}{\rho}\right)^2+\frac{\rho''}{\rho}\right)=0.\nn
\ea
Using the field equation \eqref{c7} this yields 
\be\label{G14c}
\int_z\sigma^{\frac{D_1}{2}}\rho^{\frac{D_2}{2}}
\left(\frac{2\Lambda_2}{\rho}-\frac{4}{D_1+D_2-1}\tilde V\right)=0,
\ee
which is precisely the brane constraint \eqref{T25}. 

There is a simple necessary condition for the action to be an extremum with respect to arbitrary variations of the metric and scalar field. We may first compute the most general solution to the local field equations. It will contain a number of integration constants that we denote by $\gamma_i$. We next insert the solution into the action and compute $S(\gamma_i)$. A necessary extremum condition is $\partial S/\partial\gamma_i=0$, or, with eq. \eqref{c04},
\be\label{95A}
\frac{\partial}{\partial\gamma_i}\int_z
\rho^{\frac{D_2}{2}}_\gamma
\sigma^{\frac{D_1}{2}}_\gamma
\tilde V_\gamma=0.
\ee
(We denote  by $\rho_\gamma(z)$ etc. the solution corresponding to a particular set of integration constants $\gamma$.) In particular, in absence of a scalar field and for constant $\tilde V\neq 0$, the condition \eqref{95A} implies that $\int\rho^{\frac{D_2}{2}}\sigma^{\frac{D_1}{2}}$ be at an extremum with respect to variations of the integration constants. We recall that eq. \eqref{95A} is necessary, but not sufficient for an extremum of the action.

\begin{itemize}
\item []{\bf b)\quad non-local field variations}
\end{itemize}

Our findings so far clearly indicate that for generic solutions of the field equations the action is not an extremum. While it is an extremum with respect to local field variations, it is not for nonlocal field variations. It is a crucial question if the action has to be an extremum with respect to such non-local field variations. Let us address this question from a different point of view - the ``brane point of view''. For this purpose we consider codimension two singularities in a six dimensional space $(d=6,~D_1=4,~D_2=1$). For the brane singularities, where $\rho(z)$ vanishes for $z=0$ or $z=\bar z$, we may take the point of view that a singular object - a brane - ``sits'' at the singularity. This corresponds to an additional singular contribution to the energy momentum tensor $\sim\delta(z)$ or $\sim\delta(z-\bar z)$, given by the brane tension. The brane tension is in one to one correspondence with the deficit angle such that the properties of the brane can actually be inferred from the properties of the bulk geometry \cite{SCW}. We take a similar point of view for the zerowarp singularity and use the word ``brane tension'' also for the singular energy momentum tensor for this configuration. 

Consider the solutions of pure six dimensional gravity with a cosmological constant in sect. \ref{Einsteinspaces}. They have two physical integration constants. For the solutions with two branes those can be associated to the two deficit angles at the two singularities. In other words, the two integration constants are given by the tension of the two branes. If we consider the brane tensions 
as fixed intrinsic values, the bulk metric is completely fixed. (This holds up to the two irrelevant integration constants, one giving the location of branes, i.e. the location of one brane at $z=0$, the other specifying the overall normalization of $\sigma$.) For fixed brane tensions the boundary conditions at the singularities are fixed by appropriate ``matching conditions''. The variations of the metric which lead to the brane constraint \eqref{T13} are not compatible with those fixed boundary conditions. Indeed, the metric is fully specified by the solution of the field equations with the boundary conditions given by the two brane tensions. Now $\Lambda_1$ is given by the values of the two brane tensions and there seems to be no reason anymore to impose the constraint \eqref{T13}.

This situation generalizes to the singularities with codimension larger than one. There is a one to one correspondence between the singular energy momentum tensor on the brane on one side, and physical integration constants which parameterize the most general solution of the gravitational field equations with the symmetries of our ansatz, on the other side. For two singularities the two brane tensions fix two integration constants. In the absence of a scalar field this fixes $\Lambda_1$. These considerations also hold for the case where some singularity of our solution is a coordinate singularity, like the horizon of the black hole. (The location of the ``brane'' is then inside the horizon.)

If we add the scalar field, two additional integration constants characterize the most general solution, i.e. $\delta(z_0)$ and $\delta'(z_0)$. Even for fixed brane tensions the effective $D_1$-dimensional cosmological constant $\Lambda_1$ is no longer fixed uniquely. It will depend on the additional integration constants. Thus $\Lambda_1$ appears to be again a free integration constant. Of course, we can fix the two additional integration constants by the properties of the scalar field at the singularities. In this sense, we can define a ``scalar charge'' of the branes, similar to the``cosmon charge'' of black hole like solutions in four-dimensional gravity with scalars \cite{CWBH}. Then $\Lambda_1$ is determined by the brane tensions and the scalar charges of the two branes. For fixed scalar charges, the variations of $\delta$, which lead to the boundary term in eq. (92), would then be replaced by a fixed behavior of $\delta$ at the singularity, as given by the scalar charge. 

We note that in distinction to the energy momentum tensor no conserved charge can be associated to the ``scalar charge'', even if a singularity is embedded in flat higher dimensional space. Fixing the scalar charges of the two branes is just another wording for taking fixed values of the integration constants $\delta(z_0)$ and $\delta'(z_0)$ and seems quite ad hoc. Constraining the variations of the metric and scalar field to fixed values of the scalar charges lacks any natural motivation. In particular, there is no reason why the scalar charges cannot change in the course of the cosmological evolution.

We will argue in the next section that this consideration also holds for the brane tensions, unless additional unknown microphysics keeps the brane tension fixed. Similar to the mass of a black hole in usual four-dimensional gravity, the brane tensions should be considered as dynamical quantities, reflecting the result of a given cosmological history. For a black hole, one does not consider its mass as given a priori -- it is rather a result of the accretion of mass in the past. There is no difference between a black hole and a brane in higher dimensions, except the different value of $D_1$. (For black holes one has $D_1=1$). We therefore will consider the brane tensions as dynamical quantities as well.

\section{Cosmology with varying brane tension}
\label{Cosmologywith}
For variable brane tensions we will investigate in this section the variations of the higher dimensional metric and scalar fields from a dynamical point of view. We will discuss the possible time evolution of these fields. This will shed new light on the question why a static cosmological solution is subject to the brane constraint \eqref{T13}. If the brane tensions are not fixed a priori, but are rather allowed to vary with time, there seems to be no reason to exclude from the allowed states the metric variations of $g^{(D_1)}\mn(x)$, which are local in $D_1$-dimensional space, but global in the coordinates of internal space. Computing the brane tension for some fixed internal geometry and some arbitrary given metric $g^{(D_1)}\mn(x)$ will result in a brane tension that depends on $x^\mu$. 

In analogy to the four-dimensional black holes, this seems rather natural. Indeed, for a study of the time evolution of black hole like objects the metric functions $A(r,t),B(r,t)$ of the isotropic line element $ds^2=-Bdt^2+Adr^2+r^2d^2\Omega$ depend on time and the radial coordinate $r$. (In particular, $B(r,t)$ is the analogue of an $x$-dependent warping factor $\sigma(z,x)$ or, more generally, of an $x$-dependent metric component $\tilde g\mn(z,x)$.) The mass of the black hole can be defined by the radial dependence of $B(r,t)$ at a given distance of the singularity. (At a sufficient distance of the black hole, $B(r,t)$ is directly related to the Newtonian potential.) In general, this mass will depend on time - in complete analogy to the higher dimensional brane tension depending on $x^\mu$. There would be no good reason to impose an additional constraint on $B(r,t)$ which keeps the  mass of the black hole at some given time independent value. (Note that the mass of the black hole is conserved only if it is embedded in empty flat space. If there are stars or dust outside the region which is used to define the mass of the black hole inside, these objects may be accreted later and enhance the mass of the black hole.) Similarly, there seems to be no good reason to restrict the possible metrics $g^{(D_1)}\mn$ to those with a fixed brane tension. 

All these remarks may seem rather trivial. The consequences for our understanding of singular solutions in higher dimensional space-time are far reaching, however. They result in the additional brane constraints for the allowed static solutions, as the constraint \eqref{T13}. Consider again the finite-volume solutions of higher dimensional gravity discussed in sect. \ref{Einsteinspaces}, and start at some time $t_0$ with a metric configuration given by such a solution. We will argue that for $t>t_0$ neither $L$ nor $g^{(D_1)}\mn$ remain at the values given at $t_0$, i.e. $L(t_0)=1,R^{(D_1)}\mn(t_0)=\Lambda_1g^{(D_1)}\mn$, unless the brane constraint is met. In consequence, also the brane tensions do not remain at the same values as for $t_0$ - they change in the course of the cosmological evolution. 

The action for the dimensionally reduced $D_1$-dimensional theory for the metric $g^{(D_1)}\mn$ and the dimensionless ``radius of internal space'' $L$ reads
\ba\label{93A}
&&S^{(D_1)}=\M\int_x(g^{(D_1)})^{1/2}
\left\{L^{D_2+1}\left(\frac12 R^{(D_1)}-\frac{c_V}{c_R}\right)\right.\nn\\
&&\left.+L^{D_2-1}\left(\frac{D_1+D_2+1}{D_1+D_2-1}\frac{c_V}{c_R}-\frac{D_1\Lambda_1}{2}\right)\right.
\nn\\
&&\left.-\frac{c_L}{2}L^{D_2-1}
\partial^\mu L\partial_\mu L\right\},
\ea
with $c_L$ an integral over internal space which multiplies the derivative term for an $x$-dependent radius $L(x)$. We have already seen in sect. \ref{Dimesionalreduction} that $S^{(D_1)}$ is an extremum with respect to variations of $g^{(D_1)}\mn$ and $L$ only if the constraint \eqref{T13} is obeyed. We will investigate here in more detail the solutions to the field equations for $g^{(D_1)}\mn$ and $L$ that follow from the variation of the action \eqref{93A}. For simplicity we consider only the case $D_1=4$.

We first perform a Weyl-scaling
\be\label{97}
g^{(D_1)}\mn=L^{-(D_2+1)}g\mn
\ee
such that $S^{(4)}$ becomes (with $R$ the curvature scalar built from $g\mn$)
\be\label{98}
S^{(4)}=M^2\int_xg^{1/2}
\left\{\frac R2-V-\frac{\tilde c_L}{2}\partial^\mu\ln L\partial_\mu\ln L\right\}.
\ee
(Since we will consider only variations of $g\mn$ and $L$ that are local in the four-dimensional space, we are allowed to omit boundary terms.) The action \eqref{98} corresponds to the Einstein frame where the coefficient in front of $R$ is field independent. In this frame the potential reads
\ba\label{99}
&&V(L)=V^{(D_1)}
(L)L^{-2(D_2+1)}\\
&&=\frac{c_V}{c_R}L^{-(D_2+1)}
-\left(\frac{D_2+5}{D_2+3}\frac{c_V}{c_R}-2\Lambda_1\right) L^{-(D_2+3)},\nn
\ea
with derivative
\ba\label{100}
\frac{\partial V}{\partial L}&=&-(D_2+1)
\frac{c_V}{c_R}L^{-(D_2+2)}\\
&&+(D_2+5)\frac{c_V}{c_R}
L^{-(D_2+4)}
-2(D_2+3)\Lambda_1L^{-(D_2+4)}.\nn
\ea
The extremum occurs for 
\be\label{101}
L^2_0=\frac{D_2+5}{D_2+1}-
\frac{2\Lambda_1c_R}{c_V}
\frac{D_2+3}{D_2+1}.
\ee

Only if the brane constraint \eqref{T13} is met,
\be\label{101A}
\Lambda_1=\Lambda_{1,c}=\frac{2}{D_2+3}\frac{c_V}{c_R},
\ee
the extremum occurs for $L_0=1$, as required by the $d$-dimensional solution. For all other values of $\Lambda_1$ the nonvanishing derivative $\partial V/\partial L$ will induce a change of $L(t)$ from the ``initial value'' $L(t_0)=1$ to some other value. Thus the solution of the field equation for $L$ derived from the effective four-dimensional action \eqref{98} cannot be static if we start with $L(t_0)=1$ and $\Lambda_1\neq \Lambda_{1,c}$. In other words, the true solution corresponding to the extremum of the higher dimensional action will be a non-static runaway solution for such an initial condition. We will discuss the generic properties of these runaway solutions in the next section, and end this section with a few more considerations on the possible static solutions. 

For $L_0=1$ the solution of the four-dimensional field equation
\be\label{102}
R\mn-\frac12 Rg\mn=- Vg\mn
\ee
is given by
\be\label{103}
R=4V=4\Lambda_1=\frac{8}{D_2+3}\frac{c_V}{c_R}
\ee
if the brane constraint is met. In this case we have a consistent solution with static $L(t)=1$ and $R\mn=\Lambda_{1,c}g\mn$. The second derivative $\partial^2 V/\partial L^2$ at the extremum at $L_0=1$ is negative for $c_V/c_R>0$, such that $L_0$ actually corresponds to a maximum of $V(L)$. The consistent static solution is then unstable with respect to small excitations. (The vicinity of this solution could describe an inflationary universe.) 

We may understand the properties of the effective $D_1$-dimensional scalar potential $V$ in the Einstein frame in a more general context. Let us consider a set of $D_1$-dimensional scalar fields  $\psi_u$ that correspond to local variations of the $d$-dimensional fields which vanish at the singularities. A second set of scalar fields $\varphi_z$ describes variations that do not vanish at the singularity, like the variation of the overall size of internal space $L$. The fields $\psi_u$ and $\varphi_z$ are chosen such that they vanish for a solution to the $d$-dimensional field equation. Expanding around a solution of the $d$-dimensional field equations one finds in linear order for homogeneous fields $\psi_u$ the $D_1$-dimensional field equations
\be\label{QA}
D_t\psi_u=
\frac{\partial V}{\partial \psi_u}=
Q_u(\psi_u,\varphi_z)~,~Q_u(0,0)=0.
\ee
Here $D_t$ is an appropriate operator involving time derivatives. For flat $D_1$-dimensional space one has $D_t=-\partial^2_t$, while the appropriate generalization to $D_1$-dimensional (anti-) de Sitter space makes $D_t$ vanishing in the case of maximal symmetry of the $D_1$-dimensional subspace. 

The solution of the $d$-dimensional field equations guarantees an extremum of the action with respect to local variations $\psi_u$. In consequence, $Q_u$ vanishes for $\psi_u=\varphi_z=0$. Away from the solution of the $d$-dimensional field equations the source term $Q_u$ will not vanish anymore - the fields $\psi_u\neq 0$ or $\varphi_z\neq 0$ no longer correspond to an extremum of $V$. In our case the $d$-dimensional solution $L=A_\rho=1$ corresponds to $\psi_u=0,\varphi_z=0$ while other values of $L$ and $A_\rho$ are no longer higher dimensional solutions. They amount to nonvanishing $\varphi_z$, with $Q_u(\varphi_z\neq 0)\neq 0$. They therefore do not correspond to static solutions. In other words, for $L\neq 1$ or $A_\rho\neq 1$ a non-static behavior arises due to a source term for the fields $\psi_u$, even if $L$ and $A_\rho$ themselves may have no source term for some particular values.

In contrast, for the field equations for the nonlocal fluctuations
\be\label{QB}
D_t\varphi_z=\frac{\partial V}{\partial\varphi_z}=
R_z(\psi_u,\varphi_z)
\ee
no argument is available in order to guarantee $R_z(0,0)=0$. The solution of the higher dimensional field equations only guarantees an extremum for the local fluctuations, but not for the nonlocal fluctuations $\varphi_z$. In general, one has a nonvanishing source term and the fields $\varphi_z$ do not remain constant. The changes of the scales of $D_2+1$ dimensional space, $L$, or the $D_2$-dimensional subspace, $A_\rho$, correspond to nonlocal fluctuations $\varphi_z$. The solution to the higher dimensional field equation, $\psi_u=\varphi_z=0$, is not sufficient to guarantee an extremum of $V$. The extremum condition $R_z(0,0)=0$ provides for the additional brane constraints \eqref{T13} and \eqref{T25}.

\section{Runaway solutions}
\label{runawaysolutions}
In this section we discuss the fate of cosmologies which start at $t_0$ with  a higher dimensional solution for which the brane constraint is not obeyed.  We first discuss this issue within the reduced four-dimensional theory and investigate the solutions for the field equation derived from the action \eqref{98}. For $\Lambda_1>\Lambda_{1,c}$ the potential derivative is negative for $L(t_0)=1$. The radius $L$ therefore starts to increase for $t>t_0$. Since no extremum of $V$ is encountered, we have a ``runaway solution'' where $L$ evolves to infinity as $t\to\infty$. For large enough $L$ only the term $\sim L^{-(D_2+1)}$ matters for the potential. This is a rather generic type of higher dimensional runaway solutions \cite{CWCC}. For a scalar field with canonical kinetic term,
\be\label{104}
\varphi=\tilde c^{1/2}_LM\ln L
\ee
one finds an exponential potential
\be\label{105}
V\sim M^4\exp
\left(-\tilde\alpha\frac{\varphi}{M}\right)~,~\tilde\alpha=\frac{D_2+1}{\sqrt{\tilde c_L}}.
\ee
In presence of radiation and matter this leads to the typical scaling solutions associated with quintessence \cite{CWQ}, provided $\alpha>2$. For $\alpha<2$ or in the absence of matter and radiation one has a scalar field dominated late cosmology. For both types of solutions the scalar field $\varphi$ increases $\sim\ln t$ and the scalar potential decreases $\sim t^{-2}$, at the same rate as the dominant matter or radiation. The Hubble parameter decreases $\sim t^{-1}$, such that the curvature scalar $R$ vanishes asymptotically for $t\to\infty$. (For a realistic particle physics model, however, possible   problems arise from too large variations of ``fundamental coupling constants'' \cite{CWCC}.)

The solution of the effective four-dimensional theory looks very different from the one expected for the higher dimensional solutions with $\Lambda_1>\Lambda_{1,c}$. Instead of a de Sitter solution with fixed cosmological constant $\Lambda_1$, the effective cosmological constant relaxes to zero. On the other side, for $\Lambda_1<\Lambda_{1,c}$, the potential derivative is positive for $L(t_0)=1$. For $t>t_0$ the radius decreases now towards zero. The terms $\sim L^{-(D_2+3)}$ will dominate the potential \eqref{99} and $V(L)$ will get negative. Again, the solution is quite different from the static six dimensional solution. For both cases with $\Lambda_1\neq \Lambda_{1,c}$ the boundary terms at the singularities change. These solutions correspond to cosmologies with brane tensions changing with time.

Of course, the solutions of the effective four-dimensional theory for $g^{(D_1)}\mn$ and $L$ are not solutions of the higher dimensional field equations. They obtain for a fixed shape of the internal geometry, i.e. fixed functions $\rho(z),\sigma(z)$. For true dynamical solutions, which should also obey the $d$-dimensional field equations, one expects that other fields except $g^{(D_1)}\mn$ and $L$ change their values for $t>t_0$. For example, the functions $\rho(z)$ and $\sigma(z)$ correspond to infinitely many four-dimensional scalar fields after dimensional reduction, the ``Kaluza-Klein modes''. Typically, these are massive scalar fields, often with a discrete spectrum \cite{NW}. The four-dimensional field equations discussed above neglect the changes in these massive scalar fields. 

However, one does not expect that these massive modes change the qualitative behavior of the solution. The runaway solutions for $L\to\infty$ or $L\to 0$ correspond to rather generic geometrical features. They involve only considerations of relevant length scales, rather than the detailed shape of internal geometry, reflected in the detailed properties of the scalar potential \cite{CWCC}. If one would ``integrate out'' these modes by solving their field equations as functions of given $L(t),g^{(D_1)}\mn(t)$, one would obtain a modified effective action \eqref{93A}. The potential $V(L)$ would get modified, and the constant $c_L$ as well as the coefficient of the curvature term would acquire an additional $L$-dependence. The general shape of $V(L)$ cannot change, however. Its only true extremum must correspond to a static solution of the higher dimensional field equations. They have all been classified in sect. \ref{Einsteinspaces}. Furthermore, the brane constraints must be satisfied. The only true extremum is therefore the maximum at $L_0=1$. We conclude that the only static solution corresponds to the maximum at $L_0=1$, with $\Lambda_1=\Lambda_{1,c}$. Since this solution is unstable, it requires very particular initial conditions. 

All other initial conditions end in runaway solutions with $L\to\infty$ or $L\to 0$. In particular, for initial conditions given by a higher dimensional solution with $\Lambda_1\neq\Lambda_{1,c}$, the behavior of $L$ very near the initial time $t_0$ is given by $V(L)$ in eq. \eqref{99}. All massive modes start with small values at $t_0$ as well, and they do not influence the evolution of $L$ in a linear approximation in $L-1$. Starting from a static higher dimensional solution, with arbitrary integration constants corresponding to the continuous family of solutions of ref. \cite{RDW}, at most a subset with discrete values of the four-dimensional cosmological constant really shows static behavior. For variable tension branes generic initial conditions lead to runaway solutions.

We have argued that the generic character of the runaway solutions remains even we include the other degrees of freedom obtained from dimensional reduction. Indeed, the runaway solutions correspond to the generic extrema of the higher dimensional action - they are no artefacts of the approximation \eqref{93A}. As an example for the quantitative changes induced by the presence of additional fields we include further scalar fields in the effective $D_1$-dimensional action. In fact, the issue of additional changes of internal geometry beyond the overall scale can be demonstated by including into the effective $D_1$-dimensional theory also the scalar fields that correspond to $x$-dependent factors $A_\rho(x)$ and $A_\sigma(x)$. This yields
\ba\label{V1}
S\de&=&\M\int_x(g\de)^{1/2}\nn\\
&&\left\{\frac12 w^{-2}R\de-V\de+{\cal L}\de_{kin}\right\}
\ea
with
\be\label{V2}
w^{-2}=L^{D_2+1} A^{\frac{D_2}{2}}_\rho A^{\frac{D_1-2}{2}}_\sigma
\ee
and
\ba\label{V3}
&&V\de =L^{D_2+1}A^{\frac{D_2}{2}}_\rho A^{\frac{D_1}{2}}_\sigma\left\{\frac{c_V}{c_R}
+\frac{D_1\Lambda_1}{2L^2}\right.\\
&&\left.+\frac{D_2\Lambda_2}{2L^2}\frac{c_\rho}{c_R}
\left(1-\frac{1}{A_\rho}\right)\right.
\left.-\frac{1}{L^2}
\frac{D_1+D_2+1}{D_1+D_2-1}\frac{c_V}{c_R}\right\}.\nn
\ea
We observe that $V\de$ is monotonic in $A_\sigma$ with a vanishing derivative only for $V\de=0$. At first sight this seems to question the consistency of static solutions for $V\de\neq 0$. However, for $\Lambda_1\neq 0$ the term $R\de$ will also contribute to the $A_\sigma$-dependence of the effective action, as discussed after eq. \eqref{T17}. There are now also derivative terms for $A_\rho$ and $A_\sigma$
\ba\label{V4}
{\cal L}\de_{kin}&=&-\frac12 L^{D_2+1}A^{\frac{D_2}{2}}_\rho
A^{\frac{D_1-2}{2}}_\sigma\\
&&\{c_LL^{-2}\partial^\mu L\partial_\mu L+c_{K\rho}A^{-2}_\rho\partial^\mu A_\rho\partial_\mu A_\rho\nn\\
&&+c_{K\sigma}A^{-2}_\sigma\partial^\mu A_\sigma\partial_\mu A_\sigma+ {\textup mixed ~terms}\}.\nn
\ea
The mixed terms contain terms $\sim \partial^\mu L \partial_\mu A_\rho$ etc..

We restrict the discussion again to $D_1=4$ and perform the Weyl scaling
\be\label{V5}
g^{(4)}\mn=w^2g\mn,
\ee
resulting in the effective action in the Einstein frame
\be\label{V6}
S^{(4)}=M^2\int_xg^{1/2}
\left\{\frac{1}{2}R-V+{\cal L}_{kin}\right\},
\ee
with scalar potential
\ba\label{V7}
V&=&w^4V^{(4)}=L^{-(D_2+1)}A_\rho^{-\frac{D_2}{2}}
\left\{\frac{c_V}{c_R}+\frac{2\Lambda_1}{L^2}\right.\\
&&\left.+\frac{D_2\Lambda_2}{2L^2}\frac{c_\rho}{c_R}
\left(1-\frac{1}{A_\rho}\right)-\frac{1}{L^2}
\frac{D_2+5}{D_2+3}\frac{c_V}{c_R}\right\}.\nn
\ea
After Weyl scaling the potential shows no dependence on $A_\sigma$ anymore. This simply reflects that $A_\sigma(x)$ and $g\de\mn(x)$ are not independent variables, since $A_\sigma(x)$ can be absorbed into a redefinition of $g\de\mn(x)$. The Weyl scaled metric in the Einstein frame is independent of this artificial separation. In consequence, no kinetic term for $A_\sigma$ remains after the Weyl scaling
\ba\label{V8}
{\cal L}_{kin}&=&-\frac12
\{\tilde c_L\partial^\mu\ln L\partial_\mu\ln L+\tilde c_{K\rho}\partial^\mu
\ln A_\rho\partial_\mu\ln A_\rho\nn\\
&&+\tilde c_{L\rho}\partial^\mu\ln L\partial_\mu\ln A_\rho\}.
\ea

Besides gravity the effective four-dimensional theory contains now two scalar fields, 
$\varphi_L=\tilde c_L^{1/2}M\ln L$ and $\varphi_\rho=\tilde c^{1/2}_{K\rho}M\ln A_\rho$. Static solutions require an extremum with respect to variations of both $L$ and $A_\rho$, and consistency with the $d$-dimensional field equations requires that this extremum occurs for $L=1,A_\rho=1$. In a more general setting, there will be additional four-dimensional scalar fields, which correspond to local changes of internal geometry, i.e. changes in the $z$-dependence of $\sigma(z)$ or $\rho(z)$ which vanish at the singularities. For extrema of $V$ \eqref{V7} at $L\neq 1$ or $A_\rho\neq 1$, the variations of the extended scalar potential with respect to these additional fields will not vanish. 

Let us investigate the shape of the potential \eqref{V7} more closely. With $\tilde c_V=c_V/c_R~,~\tilde c_A=c_\rho/c_R~,~\Lambda_2=D_2-1$ and $\Lambda_1\to \Lambda~,~A_\rho\to A$ this is the effective potential \eqref{A}. First of all, we note that for $\Lambda_2=0$ there is no minimum with respect to variations of $A_\rho$ unless $V=0$. This is a manifestation of the discussion after eq. \eqref{T26}. For $D_2=1$, where $\Lambda_2=0$, the only possible solution for static internal geometries with finite volume can occur for 
\be\label{V9}
\frac{c_V}{c_R}=\frac{1}{L^2}
\left\{
\frac{D_2+5}{D_2+3}\frac{c_V}{c_R}-2\Lambda_1\right\}.
\ee
On the other hand, an extremum of $V$ with respect to $L$ requires eq. \eqref{101}. For static solutions this has to occur for $L_0=1$ and therefore obeys the constraint \eqref{T13}, \eqref{101A}, $\Lambda_1=2c_V/\big[(D_2+3)c_R\big]$. Insertion into eq. \eqref{V9}, together with $L=1$, yields for $D_2=1$ the condition $c_V=0$. Static solutions are then only possible for $\tilde V=0,~c_V=0,~\Lambda_1=0$. For $D_2>1,\Lambda_2\neq 0$ the requirement of an extremum of the potential \eqref{V7} at $L=A_\rho=1$ precisely yields the two conditions \eqref{T13} and \eqref{T25}. We note that the extremum of $V$ is a maximum both with respect to variations of $L$ and $A_\rho$. The static solution is unstable.

We conclude again that the generic cosmological solutions for variable tension branes are runaway solutions where $L$ and $A_\rho$ either diverge or go to zero. Let us consider initial conditions for which both $L$ and $L^2A_\rho$ increase towards infinity for asymptotic times $t\to\infty$, while $A_\rho$ goes to zero. First of all, we note that the particular runaway cosmology is modified as compared to the case $A_\rho=1$ discussed at the beginning of this section. Nevertheless, the qualitative features remain similar. A given cosmology will now move along some trajectory in $(L,A_\rho)$-space. We can associate this trajectory with a cosmon field $\varphi$ and choose a standard normalization of its kinetic term. Then we will encounter again an exponential potential decreasing to zero as $\varphi\to\infty$, as characteristic for quintessence cosmologies. These features will remain similar if we include further scalar fields beyond $L$ and $A_\rho$. The exponential potentials are generic for runaway solutions \cite{CWCC}.

For $L\to \infty,L^2 A_\rho\to\infty$ the brane tensions at the singularities go to zero. One may wonder if cosmology approaches asymptotically a state where the difference between the $D_1-1$ large space dimensions and the $D_2+1$ internal space dimensions vanishes, such that four-dimensional gravity is no longer meaningful. This is not automatic. The issue depends on the relative size of $L$ and the $D_1$-dimensional Planck mass, as well as on the curvature of $D_1$-dimensional space. We have to compare typical masses of the Kaluza-Klein modes $M_{KK}$ with the scale of gravity. Only if $M_{KK}$ vanishes sufficiently fast as compared to the effective Planck mass the internal dimensions become asymptotically observable. (We assume here implicitely that particle physics masses as the nucleon mass scale proportional $M$, as required by observation, cf. our conclusions.)

Let us consider $D_1=4$ and compute in the Weyl frame (before the Weyl scaling) the ratio $R_{KK}$ between $\tilde M^2_{KK}$ and the coefficient in front of the four-dimensional curvature scalar $R^{(4)}$. With a renormalized mass in the Weyl frame $\tilde M^2_{KK}\sim L^{-2}$ one finds
\be\label{ZA}
R_{KK}\sim L^{-(D_2+3)}A^{-\frac{D_2}{2}}_\rho.
\ee
The dimensionless ratio $R_{KK}$ is invariant under Weyl scaling such that in the Einstein frame $M^2_{KK}=R_{KK}M^2$. For the potential \eqref{V7} we may compare $M_{KK}$ with the characteristic scale for dark energy $(M^2V)^{1/4}$ and compute the asymptotic ratio
\be\label{ZB}
M_{KK}(M^2V)^{-1/4}\sim M^{\frac12}\left(\frac{c_V}{c_R}\right)^{-\frac14}
L^{-\frac{D_2+5}{4}}A_\rho^{-\frac{D_2}{8}}.
\ee
If this ratio becomes smaller than one a description in terms of a four-dimensional quintessence cosmology ceases to be meaningful. (The presently observed dark energy scale is $(M^2V)^{1/4}\approx 2\cdot 10^{-3}eV$, and $M_{KK}$ would become smaller.) We may also make a comparison with a typical size of the Hubble parameter, $H^2\sim V^{1/2}$,
\be\label{139A}
M_{KK}V^{-1/2}\sim M\left(\frac{c_V}{c_R}\right)^{-1/2}L^{-1}.
\ee
If this ratio is of the order one the size of internal space is comparable to the horizon of the four-dimensional space. The issue depends on the evolution of $c_V/c_R$. For constant $\tilde V$ observable internal dimensions seem unavoidable in our model for large $t$. On the other hand, a runaway where $c_V/c_R$ approaches zero sufficiently fast may be compatible with unobservable internal dimensions. Realistic cosmologies should correspond to increasing ratios \eqref{ZB}, \eqref{139A} for increasing $t$. We also observe that for $A_\rho\gg 1$ the lowest masses of the Kaluza-Klein modes pick up an additional factor $A^{-1/2}_\rho$ multiplying eqs. \eqref{ZB}, \eqref{139A}. In ref. \cite{CWCC} we have discussed various scenarios for runaway solutions where internal space remains unobservable and the dangerous change of ``fundamental couplings'' may be effectively stopped. 

We end this section with a brief discussion of the runaway solutions for higher dimensional theories with a scalar field. This can be done on the basis of the effective four-dimensional action \eqref{95B}. For $\tilde V>0$ the effective potential $V_\delta$ is positive definite. For the exponential potential \eqref{s1} one finds that $V_\delta$ vanishes only if $\delta\to\infty,\delta'\to 0$. For this case one may argue that it is likely that the runaway solution leads to $\delta$ increasing to infinity, with $\delta'^2$ relaxing to a small value. The runaway solution ``turns off'' the scalar field. In other words, one expects that the ratios $c_V/c_R$ and $c_K/c_R$ go to zero during the runaway. 

\section{Conclusions and discussion}
\label{X}
We have distinguished between fixed tension branes and variable tension branes. For the variable tension branes we have identified additional constraints besides the requirement of being a solution of the field equations. These ``brane constraints'' result from the extremum condition for the action with respect to field variations that change the strength of the singularity and therefore the brane tension. A simple example for such  variations are changes of the metric that are local in four-dimensional space, but do not change the geometry of internal space and the warping in a higher dimensional context. The brane constraints are also needed in order to obtain a consistent dimensionally reduced effective theory in four dimensions, where the metric can be varied without constraints.

The brane constraints strongly restrict the number of static solutions or, more generally, of solutions with maximal four-dimensional symmetry. (We assume homogeneity and isotropy in the large three space dimensions as appropriate for cosmological solutions.) While the $d$-dimensional field equations admit continuous families of solutions with free integration constraints, these are reduced to at most a discrete subset of extrema of the action. The four-dimensional cosmological constant remains no longer a free integration constant of the solution. For the static solutions it becomes fixed by the parameters of the theory, similar to the case of fixed tension branes. As a result, the singular spaces with branes become in this respect very similar to regular or compact spaces. Both have at most a discrete set of static solutions. 

We have investigated $d$-dimensional gravity with a cosmological constant. Furthermore, we have also considered additional $d$-dimensional scalar fields. For solutions with variable tension branes and finite four-dimensional gravitational constant we have found a constraint for the effective four-dimensional cosmological constant $\Lambda_1$. In particular, for flat four-dimensional space, $\Lambda_1=0$, the $d$-dimensional cosmological constant must vanish, $\tilde V=0$, cf. eq. \eqref{T13}. (For higher dimensional scalar fields this generalizes to a weighted integral over the scalar potential, $c_V=0$, cf. eq. \eqref{71A}.) A further constraint for internal geometry requires for $\Lambda_1=0$ that also the effective cosmological constant for the internal $d-5$ dimensional hypersurface must vanish, $\Lambda_2=0$. The static solutions with $\Lambda_1=\Lambda_2=\tilde V=0$ are known, cf. the appendix. They correspond either to flat $d$-dimensional space or to exact infinitely spread brane or zerowarp solutions with a powerlike decrease or increase of the warp factor for $z\to\infty$, cf. eq.\eqref{H1}. These solutions have no finite four-dimensional gravitational constant. For our class of models no variable tension branes with static solutions and effective four-dimensional gravity with vanishing cosmological constant exist. Solutions of this type that have been considered previously in the literature \cite{RSW,RDW} cannot be realized for variable tension branes.

Our general discussion can be extended to models where static solutions with $\Lambda_1=0$ and effective four-dimensional gravity are known to exist. In fact, every solution of the higher-dimensional field equations with a regular space automatically obeys the brane constraints. In the absence of singularities these constraints follow from the solution of the field equations since the boundary terms in the variation of the action vanish. It is an interesting open question whether there are also singular static spaces that obey the brane constraints, have a finite four-dimensional gravitational constant and a vanishing four-dimensional cosmological constant.

The absence of static solutions, both for $\Lambda_1\neq 0$ and vanishing $\Lambda_1$, may offer interesting perspectives for cosmology. In fact, for a positive $d$-dimensional cosmological constant $\tilde V>0$ we have found that our model does not have any stable static solution with effective four-dimensional gravity and maximal symmetry. Thus an asymptotic state of the universe with a fixed nonzero cosmological constant is excluded for the variable tension branes in our models. We argue that the characteristic cosmological solutions for variable tension branes are runaway solutions where the effective four-dimensional cosmological constant or dark energy approaches zero as time goes to infinity. These are candidates for models of quintessence. We have shown that variable tension branes typically lead to an exponentially decaying potential for the cosmon field. However, for the runaway solutions also the length scales of internal space are evolving. This poses problems with the observed approximative constancy of the couplings of the standard model of particle physics. What will be needed, is a stabilization mechanism for the ``fundamental couplings''. Several possible mechanisms have been suggested in \cite{CWCC}. We do not address this issue in the present paper.

We may compare our findings for variable tension branes with the situation for fixed tension branes. Now (part of) the brane constraints are absent, at the price of introducing the brane tension as an additional fixed parameter of the model. Performing dimensional reduction as in sect. \ref{Einsteinspaces} one may naively conclude that no consistent dimensionally reduced four-dimensional gravity theory exists even for this case. This raises the question how the concept of fixed tension branes can be implemented in higher dimensions in a meaningful way. The problem is that independent local variations of the four-dimensional metric, leaving internal geometry fixed, are no longer allowed. Since they change the strength of the singularity locally, one has to demand that for a fixed brane tension such metric variations are always accompanied by local changes in the internal geometry which compensate their effect on the brane tension. One needs to implement the constraint of a fixed brane tension within the effective four-dimensional theory. This amounts to a complicated non-linear constraint among the infinitely many four-dimensional fields that result from an expansion of the higher dimensional fields in some complete system of functions. In other words, a consistent ansatz for the four-dimensional metric is no longer of the form of eq. \eqref{A1} with arbitrary $g^{(4)}\mn(x)$, but all other fields independent of $x^\mu$. For a given $g^{(4)}\mn(x)$ also the functions $\sigma$ and $\rho$ (and possibly other functions) will have a specific dependence on $x^\mu$. Only in this way an effective four-dimensional theory can be obtained which is consistent with a fixed brane tension. While dimensional reduction for variable tension branes is straightforward, it becomes rather involved for fixed tension branes.

Let us finally address the cosmological constant problem in the light of our findings. We have encountered two different types of solutions that are extrema of the higher dimensional action: static solutions and runaway solutions. Static solutions have the maximal four-dimensional symmetry of Minkowski-, de Sitter- or anti-de Sitter-space. They occur both for fixed and variable tension branes. The solutions of the higher dimensional field equations are subject to additional constraints, either from the fixed brane tension or from the brane constraint that variable tension branes must obey. Typically, for the static solutions the effective four-dimensional cosmological constant has a value which is fixed by the parameters of the theory. Obtaining a zero or very small value of the cosmological constant requires an apparent fine tuning of parameters. In other words, such values need some additional explanation. On the other hand, the dimensionless couplings of the standard model of particle physics have naturally static values, such that no problems with a too strong time variation of those couplings arises.

For the runaway solutions, in contrast, the effective four-dimensional cosmological ``constant'' typically goes to zero as time increases to infinity. This asymptotically vanishing value is a simple consequence of geometry and therefore stable with respect to quantum fluctuations. This may solve the problem of the cosmological constant by a quintessence type cosmological scaling solution. There the value of the dark energy density is naturally of the same order as the energy density in matter and radiation - both decrease with the inverse second power of time. Indeed, for the runaway solutions the effective potential for the scalar cosmon field turns out to be of the exponential type which leads to such scaling solutions as cosmological attractors. In order to explain the recent relative increase of the dark energy fraction (``why no problem'') the scaling solutions need some additional ingredient which causes an end of the scaling regime - one possible mechanism could be an effective stop of the time evolution of the cosmon field triggered by a growing neutrino mass \cite{CWN}. The potential problems from this scenario arise from a different side. Since the geometry of internal space and the warping are not static, the rather strict bounds on a possible time variation of fundamental couplings in particle physics have to be explained.

Again, runaway solutions can exist both for fixed tension and variable tension branes. In a certain sense, they are more generic for variable tension branes, since the stabilizing effect of a fixed brane tension is absent. In our specific model the characteristic cosmological solutions are all runaway solutions, except for a few unstable static solutions for singularities with codimension larger than two. Variable tension branes offer perhaps also more freedom for a solution of the problem of ``running couplings'' in runaway solutions. It is conceivable that the adjustment of the strength of the singularity during the runaway occurs in a way that leaves the dimensionless couplings of elementary particle physics fixed. 

Within our class of models, we have found an interesting class of runaway solutions, for which the asymptotic behavior approaches the static solution with $\tilde V\to 0,~\Lambda_1\to 0,~\Lambda_2/\rho\to 0$. This static solution may not be flat $d$-dimensional space, but rather a brane or zerowarp solution with $\rho\sim z^\gamma,\sigma\sim z^{-\eta}$ for all $z$. For the runaway solution the coefficient $c_R$, which multiplies the four-dimensional curvature scalar $R^{(4)}$, may remain finite, diverging only for $t\to\infty$. In this case an effective four-dimensional gravity may be well defined. For the models with a higher dimensional field the limit $\tilde V\to 0$ corresponds to a free massless scalar field. This is the Weyl-scaled version of a higher dimensional model with dilatation symmetry \cite{CWCC}. In the asymptotic limit the brane or zerowarp solution therefore corresponds to the spontaneous breaking of an exact dilatation symmetry, with an exactly massless cosmon field as the corresponding Goldstone boson \cite{CWCC}. 

The effective four-dimensional theory for our ansatz with $SO(D_2+1)$ isometry will contain the gauge bosons of a non-abelian Yang-Mills theory with gauge symmetry $SO(D_2+1)$. The gauge coupling $g$ is dimensionless in four dimensions - it mimics the dimensionless couplings of the standard model of elementary particle physics. We may ask if it is possible for a runaway solution that $g$ runs towards a nonvanishing fixed point value $g_*$ as time increases to infinity. Then one may expect that for late times the change of $g$ is small. In a more realistic analogue this could explain the observed (approximate) constancy of the couplings. The four-dimensional non-abelian gauge theory leads to a coupling depending on the length scale, inducing a confinement scale $\tilde\Lambda$ by dimensional transmutation. If we identify $g$ with the ``running coupling'' $g(M)$ evaluated at the Planck scale, a fixed $g$ would result in a fixed (exponentially small) ratio $\tilde\Lambda/M$. This could be the analogue for a fixed ratio between the nucleon mass and the Planck mass in a more realistic scenario. 

For a realistic runaway solution we require the following conditions:
\begin{itemize}
\item [i)] The effective cosmological constant should go asymptotically towards zero. A characteristic value for the present time which is compatible with observation is $V/M^2\approx 10^{-120}$.
\item [ii)] The dimensionless gauge coupling $g$ should asymptotically approach a constant value.
\item [iii)] The mass of the Kaluza-Klein excitations should remain sufficiently large such that internal space has not yet been observed. For the present time this requires $M_{KK}/M\gtrsim 10^{-14}$. 
\end{itemize}

We argue that a runaway solution is not incompatible with a constant $g$. By simple scaling arguments one finds
\be\label{Z19a}
g^{-2}\sim L^{D_2-1}A_\rho^{\frac{D_2-2}{2}}.
\ee
Despite an increasing $L$ during a runaway solution one may have a simultaneous decrease of $A_\rho$ towards zero such that $g$ remains constant. This would require a behavior during the runaway where
\be\label{Z19b}
A_\rho\sim L^{-\frac{2D_2-2}{D_2-2}}.
\ee
The effective radius of the $D_2$-dimensional subspace $\sim L\sqrt{A_\rho}\sim (g^2L)^{-\frac{1}{D_2-2}}$ would decrease for a solution with increasing $L$ and constant $g$. On the other hand, the effective four-dimensional scalar potential (or ``cosmologocal constant'') scales
\ba\label{Z19c}
\frac{V}{M^4}&\sim& L^{-(D_2+1)}A_\rho^{-\frac{D_2}{2}}\tilde c_V\nn\\
&\sim& g^{\frac{2D_2}{D_2-2}}L^{\frac{2}{D_2-2}}\tilde c_V.
\ea
A fixed $g$ does  not contradict $V/M^4\to 0$, which may be due to $\tilde c_V\to 0$. Thus a runaway of the cosmological constant towards zero could be compatible with a fixed gauge coupling. Also internal space could remain unobservable, requiring $R_{KK}\gtrsim 10^{-28}$. The ratio $R_{KK}=M^2_{KK}/M^2$ behaves for fixed $g$ as $L^{-\frac{2(D_2-3)}{D_2-2}}$. For a realistic model with a moderate increase of $L$, a decrease of $\tilde c_V$ with a sufficiently high power of $L^{-1}$ would be needed for an acceptable present dark energy density. Another option may be decreasing $L$, with $A_\rho$ increasing according to eq. \eqref{Z19b}. Then $R_{KK}$ picks up an additional factor $A^{-1}_\rho$, such that $R_{KK}\sim L^{\frac{4}{D_2-2}}$. 

A particular simple from  of runaway would only rely on decreasing $\tilde c_V$, while $L$ and $A_\rho$ remain approximately constant. Quite generally, one finds for $D_1=4$
\be\label{149}
g^{-2}\sim \int_z\rho^{\frac{D_2-2}{2}}
\ee
such that a constant gauge coupling obtains for a static internal geometry or, more generally, if the function $z\rho^{\frac{D_2-2}{2}}$ remains essentially constant around its maximum. The gauge coupling is independent of the warping $\sigma$, however, and therefore not affected by a possible runaway of the warping near the singularities. Nevertheless, the warp factor influences the masses of the Kaluza-Klein modes and its change is restricted from this side. If realistic phenomenology requires within our model an almost static internal geometry, a possible runaway could only concern the scalar sector, with a runaway of the scalar potential towards zero. It remains to be seen if a runaway of this type can be realized in an appropriate model. A search for such possibilities opens interesting new perspectives for a solution of the cosmological constant problem. 

\medskip\noindent
{\bf Acknowledgment}\\
The author thanks the KITP, Santa Barbara, for hospitality. This research was supported in part by the National Science Foundation under Grant No. PHY05-51164.

\section*{APPENDIX: FLAT BRANES IN HIGHER DIMENSIONAL GRAVITY WITHOUT COSMOLOGICAL CONSTANT}
\renewcommand{\theequation}{A.\arabic{equation}}
\setcounter{equation}{0}
An interesting special case is $d$-dimensional gravity without a $d$-dimensional cosmological constant, $\tilde V=0$, and with vanishing $D_1$-dimensional cosmological constant, $\Lambda_1=0$. This will appear as the limiting behavior for small $\tilde V$ and $\Lambda_1$. In particular, the solutions will comprise the flat branes embedded in flat space. We first consider the case of positive $\Lambda_2$ and take $\Lambda_2=D_2-1$, such that flat space is realized asymptotically for $z\to\infty$ if $\rho=z^2,~\sigma=const.$. If a brane sits at $\bar z$ its codimension is $D_2+1$. We will concentrate the discussion here to branes with codimension larger than two, $D_2\geq 2$. 

For $\tilde V=0,\Lambda_1=0,\Lambda_2=D_2-1$ eqs. \eqref{c6}, \eqref{c7} simplify
\ba\label{B1}
U'&=&-\frac{D_1}{2}U^2-\frac{D_2}{2}UW,\\
W'&=&-\frac{D_2}{2}W^2-\frac{D_1}{2}UW+\frac{2(D_2-1)}{\rho},\label{B2}
\ea
with eq. \eqref{c5} yielding
\be\label{B3}
W=-\frac{D_1}{D_2-1}U+\epsilon\sqrt{\frac{D_1(D_1+D_2-1)}{D_2(D_2-1)^2}U^2+\frac 4\rho}.
\ee
Eq. \eqref{B3} can also be written in the form
\be\label{B3X}
\frac{D_2-1}{\rho}=\frac{D_2-1}{4}W^2+\frac{D_1}{2}UW+\frac{D_1(D_1-1)}{4D_2}U^2,
\ee
such that eq. \eqref{B2} becomes
\be\label{B3Y}
W'=-\frac12 W^2+\frac{D_1}{2}UW+\frac{D_1(D_1-1)}{2D_2}U^2.
\ee

For the special case of black holes with $D_1=1$ the equations can be further simplified. With 
\be\label{B3A}
W=\frac{U}{D_2-1}\left\{-1+\epsilon
\sqrt{1+\frac{4(D_2-1)^2}{\rho U^2}}\right\}
\ee
we obtain the relation
\be\label{B3B}
(D_2-1)W^2+2UW=\frac{4(D_2-1)}{\rho}.
\ee
This can be used to write eq. \eqref{B2} in the form
\be\label{B3C}
W'=-\frac12 W^2+\frac12 UW.
\ee
As a particular case we note that for the five-dimensional black hole, $D_1=1,~D_2=3$, one has
\be\label{B3D}
U'+W'=-\frac12 (U+W)^2~,~U+W=\frac 2z,
\ee
and therefore
\be\label{B3E}
\rho=\frac{z^2}{c\sigma}.
\ee

We first concentrate on the solutions describing a flat brane embedded in $d$-dimensional flat space. The embedding into an asymptotically flat space implies the boundary condition
\be\label{B4}
\lim_{z\to\infty}\rho=z^2~,~\lim_{z\to\infty}\sigma=1~,~\lim_{z\to\infty}W=\frac 2z.
\ee
Eq. \eqref{B3} is obeyed asymptotically for $\epsilon=+1$, if $U$ decays faster than $z^{-1}$. In this case the asymptotic behavior of eq. \eqref{B2}
\be\label{B5}
W'=-\frac{D_2}{2}W^2+\frac{2(D_2-1)}{z^2}
\ee
is obeyed by eq. \eqref{B4}. The leading behavior for $U$ is given by eq. \eqref{B1}
\be\label{B6}
\frac{U'}{U}=-\frac{D_2}{z}~,~U=u_\infty z^{-D_2},
\ee
and therefore for $z\to\infty$
\be\label{B7}
\sigma=1-\frac{u_\infty}{D_2-1}z^{-(D_2-1)}.
\ee

We may fix the initial values for $\rho,\sigma$ and $U$ for some very large $z_0$, i.e. $\rho_0=z^2_0,~U_0=u_\infty z^{-D_2}_0$ and $\sigma_0,~W_0$ given by eqs. \eqref{B7}, \eqref{B3}. The fate of the solution for smaller values of $z$ depends on the integration constant $u_\infty$. For $u_\infty=0$ we recover flat $D_1+D_2+1$-dimensional space, i.e. the exact solution $\rho=z^2,\sigma=1$. For $u_\infty\neq 0$ we may have either a singularity or finite nonzero values of $\rho$ and $\sigma$ for all finite $z$. However, finite nonzero values for $\rho$ and $\sigma$ for $z\to-\infty$ are excluded. This would imply vanishing $U$ and $W$ in this limit and contradict eq. \eqref{B2}. We conclude that either $\rho$ or $\sigma$ have to vanish or diverge at some value $\bar z$ or for $z\to-\infty$. 

We may insert eq. \eqref{B3} into eq. \eqref{B1}
\ba\label{B8}
&&U'=\frac{U^2}{2D_1(D_2-1)}
\Bigg\{1-\epsilon
\sqrt{\frac{D_2(D_1+D_2-1)}{D_1}}\nn\\
&&\qquad \sqrt{1+\frac{4D_2(D_2-1)^2}{D_1(D_1+D_2-1)\rho U^2}}
\Bigg\}.
\ea
For the case of higher dimensional black holes, where $D_1=1$, this equation reads
\be\label{B9}
U'=\frac{U^2}{2(D_2-1)}
\left\{1-\epsilon D_2
\sqrt{1+\frac{4(D_2-1)^2}{\rho U^2}}\right\}.
\ee
The qualitative behavior of the solution can be understood by comparing the relative importance of the two terms under the square root in eqs. \eqref{B8}, \eqref{B9}. For large $z$ one has $\rho U^2\to u^2_\infty z^{-2(D_2-1)}$ such that the second term in the square root $\sim (\rho U^2)^{-1}$ is large. As long as $W+2U'/U$ remains negative, the importance of this term decreases a as $z$ gets lower. With
\be\label{B10}
W+\frac{2U'}{U}=-\epsilon
\sqrt{\frac{D_1(D_1+D_2-1)}{D_2}U^2+\frac{4(D_2-1)^2}{\rho}}
\ee
and $\epsilon=1$ this is indeed the case. We note that the square root multiplying $\epsilon$ never vanishes, such that by continuity $\epsilon=1$ for the whole range of $z$ where the solution exists. The solution therefore makes a smooth transition from the regime of small $\rho U^2$ for large $z$ to the regime of large $\rho U^2$ for smaller $z$.

In the second regime we can expand the square root in powers of $(\rho U^2)^{-1}$. The leading terms read for $D_1>1$
\be\label{B11}
W=\frac{D_1 U}{D_2-1}\left\{
\xi\sqrt{\frac{(D_1+D_2-1)}{D_1D_2}}-1
\right\}
\ee
and
\be\label{B12}
U'=-\frac{D_1 U^2}{2(D_2-1)}
\left\{\xi
\sqrt{\frac{D_2(D_1+D_2-1)}{D_1}}-1\right\}=
\frac{U^2}{\eta}.
\ee
Here we define $\xi=sign(U)$ and 
\be\label{B13}
\eta=-\frac{2}{D_1+D_2}
\left\{1+\xi
\sqrt{\frac{D_2(D_1+D_2-1)}{D_1}}
\right\}
\ee
corresponds to $\eta_-$ for $\xi=1$, while for $\xi=-1$ we find $\eta_+$ (cf. eq. \eqref{H2}). We note that eqs. \eqref{B12}, \eqref{B13} are also the leading contribution for $D_1=1$. The general solution 
\be\label{B14}
U=-\frac{\eta}{z-\bar z}~,~\sigma=\sigma_0(z-\bar z)^{-\eta}
\ee
shows a singularity at $\bar z$ where $\sigma$ vanishes. (Without loss of generality we can choose $\bar z=0$.) At this singularity one has
\be\label{B15}
W=\frac{\gamma}{z-\bar z}~,~\rho=\rho_0(z-\bar z)^\gamma,
\ee
and we find from eq. \eqref{B11} for $D_1>1$
\ba\label{B16}
\gamma&=&-\frac{D_1\eta}{D_2-1}
\left\{\xi
\sqrt{\frac{D_1+D_2-1}{D_1D_2}}-1\right\}\nn\\
&=&-\frac{2}{(D_1+D_2)}
\left\{\xi
\sqrt{\frac{D_1(D_1+D_2-1)}{D_2}}-1
\right\}\nn\\
&=&\frac{2+D_1\eta}{D_2}.\label{B16A}
\ea
For $\xi=1$ one has $\gamma_-$ which is negative for all $D_1>1$. The function $\rho$ diverges as the singularity is approached. Since the warp factor goes to zero $(\eta=\eta_-<0)$ we encounter a zerowarp. The contribution to volume factor including the warping goes to zero
\be\label{B15A}
\sigma^{\frac{D_1}{2}}\rho^{\frac{D_2}{2}}\sim (z-\bar z)^{(D_2\gamma-D_1\eta)/2}\sim z-\bar z.
\ee
On the other hand, for $U<0,\xi=-1$ the singular exponents are given by the positive values $(\eta_+,\gamma_+)$. Now $\rho$ goes to zero for $D_1>1$ and the solution describes the singular geometry of a brane with codimension $D_2+1\geq 3$. Since $\rho$ vanishes at the singularity, there is no horizon away from the singularity. While $R$ and $R\hmn R\imn$ remain zero as the singularity is approached, we expect that $R_{\hat\mu\hat\nu\hat\sigma\hat\lambda}R^{\hat\mu\hat\nu\hat\sigma\hat\lambda}$ and other invariants diverge, such that we have a true singularity. In the brane picture, this singularity in the metric will be produced by a singular energy momentum tensor for the brane sitting at $\bar z$. Numerical solutions yield either a brane or a zerowarp singularity, depending on the integration constants. Once we have fixed the location of the singularity at $z=0$ and the normalization $\sigma(z\to\infty)\to 1$, there remains only one free integration constant. This may be identified with the strength of the singularity, similar to the mass of a black hole, and is reflected in $u_\infty$.  

The case of black holes for $D_1=1$ is special. For $D_1\to 1$ one finds from eq. \eqref{B16A} $\eta\to-2~,~\gamma\to 0.$
This is an indication for the qualitative changes for the black hole for $D_1=1$. Indeed, for $D_1=1$ the leading behavior of $W$ for large $\rho U^2$ becomes
\be\label{B17}
W=\frac{2(D_2-1)}{\rho U}~,~U=\frac{2}{z-\bar z}
\ee
or
\ba\label{B18}
\rho'&=&(D_2-1)(z-\bar z),\nn\\
\rho&=&\rho_s+\frac{D_2-1}{2}(z-\bar z)^2.
\ea
For $\rho_s>0$ one finds close to the singularity $\rho\to const.,~\sigma\to \sigma_0(z-\bar z)^2$, such that $\rho U^2\to 4\rho_s/(z-\bar z)^2$ diverges. On the other hand, for $\rho_s=0$ both $\sigma$ and $\rho$ vanish $\sim (z-\bar z)^2$, and therefore $\rho U^2\to 2(D_2-1)$. The approximation leading to eq. \eqref{B17} does therefore not remain valid for $\rho_s=0$. 

The case $\rho_s=0$ corresponds to $\gamma=2$ for the general ansatz \eqref{H1}. In this case eq. \eqref{H2} is no longer valid. For $\gamma=2$ the terms $\sim \rho^{-1}$ matter and eqs. \eqref{B1},\eqref{B2} imply the relations
\ba\label{B22}
&&\eta=\frac{2(D_2-1)}{D_1},\\
&&2(D_2-1)-D_1\eta=\frac{2(D_2-1)}{\rho_0}.\label{B23}
\ea
Eqs. \eqref{B22} and \eqref{B23} have no solution for $D_2>1$ and finite $\rho_s$. Indeed, a numerical solution of the differential equations shows that $\rho(z\to\bar z)$ always approaches a positive constant $\rho_s>0$, for all integration constants $u_\infty\neq 0$. This shows that isotropic black holes in arbitrary dimensions always have a horizon. 

In order to make contact with the Schwarzschild metric $ds^2=-B(r)dt^2+A(r)dr^2+r^2d^2\Omega$ we identify 
\ba\label{B24}
&&r^2=\rho(z)~,~B(r)=\sigma(z)~,~A^{1/2}(r)dr=dz,\nn\\
&&A(r)=\frac{4\rho}{\rho'^2}=\frac{4}{\rho W^2}. 
\ea
The nonzero value of $\rho_s$ for $D_1=1$ implies that the singularity occurs for a nonzero Schwarzschild radius $r_s=\sqrt{\rho_s}$, corresponding to the horizon. In our coordinates the relation $A(r)B(r)=const.$ translates to
\be\label{B25}
\frac{\rho'}{\rho}+\frac{\sigma'}{\sigma}-\frac{2\rho''}{\rho'}=0.
\ee
This can be written as $W'=\frac12 UW-\frac12 W^2$ and coincides with eq. \eqref{B3C}. Using eq. \eqref{B24} our solution for black holes in arbitrary dimensions can easily be transformed to the Schwarzschild coordinates.

One may ask if a horizon can also exist for branes with $D_1>1$. A finite $\rho_s$ at the singularity corresponds to $\gamma=0$. From eq. \eqref{B16A} we conclude that this is not possible for $D_1>1$. The generalized branes embedded in flat space have no horizon, except for the black holes with $D_1=1$.

For $\tilde V=0,~\Lambda_1=0,~\Lambda_2>0$ the solutions of branes embedded in flat space are the generic solutions. We can take $\epsilon=1$ and use the fact that $\epsilon$ cannot switch since $F$ is negative (cf. eqs. \eqref{c4}, \eqref{49A}). We therefore can have only one singularity. This must be encountered for decreasing $z$. Taking the location of the singularity at $z=0$ and following the solution for positive $z$ one finds flat space for $z\to \infty$. 

The situation changes drastically for $\Lambda_2<0$. Now the sign $\epsilon$ may switch. We find indeed solutions with two singularities, one brane $(\eta_+,\gamma_+)$ and one zerowarp $(\eta_-,\gamma_-)$. The volume as well as the effective $D_1$-dimensional gravitational constant $\sim c_R$ are finite. In contrast to the solutions for $\Lambda_2>0$ an effective $D_1$-dimensional gravity exists for $\Lambda_2<0$. 

Finally, we can consider the boundary case $\tilde V=0,\Lambda_1=0,\Lambda_2=0$. In this case we find that the singular solutions \eqref{H1} , both for $(\eta_+,\gamma_+)$ and $(\eta_-,\gamma_-)$, are exact. Also $\rho(z)=\rho_0,\sigma(z)=\sigma_0$ is an exact solution. This describes flat space in cartesian or torus coordinates if we take $g\dz\ab=\delta\ab$. For all these solutions $c_R$ diverges. For $\Lambda_1=\Lambda_2=0$ the overall scaling of $\sigma$ and $\rho$ does not enter the equations such that $\sigma_0$ and $\rho_0$ are integration constants that may be set to arbitrary values without affecting the character of the solution. At some $z_{in}$ we may take $\sigma_{in}=\sigma_0 z^{-\eta}_{in},~\rho_{in}=\rho_0z^\gamma_{in}$. For an arbitrary value of the third integration constant $U_{in}\neq 0$ we can always find $z_{in}$ such that $U_{in}=-\eta_\pm/z_{in}$. This means that these initial values correspond to the exact solution \eqref{H1} for appropriate $\sigma_0$ and $\rho_0$. For $U_{in}=0$ one has the flat solution. The exact solutions \eqref{c16a} are therefore the most general solutions for $\tilde V=\Lambda_1=\Lambda_2=0$, with integration constants given by $\sigma_0,\rho_0$ and the location of the singularity $\bar z$. For $D_2=1$ only the zerowarp solution exists besides flat space.


\newpage
\begin{thebibliography}{100}
\bibitem{ARS}K. Akama, in ``Gauge Theory and Gravitation'', Proceedings Nara 1982, eds. K. Kikkawa, N. Nakanishi, H. Nairai, 267 (Springer Verlag) 1983; arXiv: hep-th/0001113;\\
V. Rubakov, M. Shaposhnikov, Phys. Lett. {\bf B125}, 136 (1983) 
\bibitem{CWWB}C. Wetterich, Nucl. Phys. {\bf B253}, 366 (1985)
\bibitem{PR}J. Polchinski, Phys. Rev. Lett. {\bf 75}, 4724 (1995);\\
L. Randall, R. Sundrum, Phys. Rev. Lett. {\bf 83}, 3370 (1999)
\bibitem{RSW}V. Rubakov, S. Shaposhnikov, Phys. Lett. {\bf B125}, 139 (1983) 
\bibitem{RDW}S. Randjbar-Daemi, C. Wetterich, Phys. Lett. {\bf B166}, 65 (1986) 
\bibitem{RS}L. Randall, R. Sundrum, Phys. Rev. Lett. {\bf 83}, 4690 (1999)
\bibitem{GS}T. Ghergetta, M. E. Shaposhnikov, Phys. Rev. Lett. {\bf 85}, 240  (2000) 
\bibitem{CS}M. Giovannini, H. Meyer, M. E. Shaposhnikov, Nucl. Phys. {\bf B619} (2001) 615; 
O. Corradini, et al., Phys. Lett. {\bf B521}, 96 (2001); J.~M. Cline, J.~Descheneau, M.~Giovannini, J.~Vinet, JHEP {\bf 0306}, 048 (2003); 
J.~Vinet, J.~M. Cline, Phys. Rev. {\bf D70}, 083514 (2004); 
I.~Navarro, Class. Quant. Grav. {\bf 20}, 3603 (2003); J.~Garriga, M.~Porrati, JHEP {\bf 0408}, 028 (2004); A.~Kehagias, Phys. Lett. {\bf B600}, 133 (2004); P.~Bostock, R.~Gregory, I.~Navarro, J.~Santiago, Phys. Rev. Lett. {\bf 92}, 221601 (2004); H.~M.~Lee, Phys. Lett. {\bf B587}, 117 (2004); Y.~Aghababaie, C.~P.~Burgess, S.~L.~Parameswaran, F.~Quevedo, Nucl. Phys. {\bf B680}, 389 (2004); H.~P.~Nilles, A.~Papazoglou, G.~Tasinato, Nucl. Phys. {\bf B677}, 45 (2004); M.~L.~Graesser, J.~E.~Kile, P.~Wang, Phys. Rev. {\bf D70}, 024008 (2004); I.~Navarro, J.~Santiago, JHEP {\bf 0502}, 007 (2005); H.~M.~Lee, A.~Papazoglou, Nucl. Phys. {\bf B705}, 152 (2005); P.~Wang, X.~H.~Meng, Phys. Rev. {\bf D71}, 024023 (2005)
\bibitem{9A}S. M. Carroll, M. M. Guica, arXiv: hep-th/0302067;\\
I. Navarro, JCAP 0309, 004 (2003)
\bibitem{SCW}J. Schwindt, C. Wetterich, Phys. Lett. {\bf B578}, 409 (2004) 
\bibitem{7A}C. Wetterich, Nucl. Phys. {\bf B253}, 366 (1985)
\bibitem{CWQ}C. Wetterich, Nucl. Phys. {\bf B302} (1988) 668
\bibitem{CWCC}C. Wetterich, Phys. Rev. {\bf D77}: 103505 (2008)
\bibitem{RSS}S. Randjbar-Daemi, A.~Salam, J.~Strathdee, Nucl. Phys. {\bf B214}, 491 (1983)
\bibitem{SWCOS}J. Schwindt, C. Wetterich, Phys. Lett. {\bf B628}, 189 (2005)
\bibitem{CWBH}C. Wetterich, Phys. Lett. {\bf B522}, 5 (2001); \\
M. Neubert, T. Krueger, C. Wetterich, Phys. Lett. {\bf B663}, 21 (2008)
\bibitem{NW}H. Nicolai, C. Wetterich, Phys. Lett. {\bf B150}, 347 (1985)
\bibitem{CWN}L. Amendola, M. Baldi, C. Wetterich, arXiv: 0706.3064 (2007); 
C. Wetterich, Phys. Lett. {\bf B655}, 201 (2007)
\end{thebibliography}
\end{document}